\documentclass[runningheads,envcountsame]{llncs}

\usepackage{amsmath,amsthm,amssymb}
\usepackage{url}

\newcommand{\comp}{\operatorname{comp}}
\newcommand{\atom}{\operatorname{atom}} 
\newcommand{\dom}{\operatorname{dom}}
\newcommand{\cdom}{\operatorname{codom}}
\newcommand{\mgu}{\operatorname{mgu}}
\newcommand{\gnd}{\operatorname{ground}}
\newcommand{\sfac}{\operatorname{sfac}}

\newcommand{\disjun}{\uplus}

\newcommand{\tenum}[2]{#1_1,\ldots,#1_#2}
\newcommand{\shortrules}[6]{\noindent\begin{minipage}{#6ex}{\bfseries #1}\end{minipage} $\;$ #2 $\;\Rightarrow_{\text{#5}}\;$ #3 \par\smallskip\noindent #4}

\begin{document}

\title{SCL(FOL) Can Simulate Non-Redundant Superposition Clause Learning}
\author{Martin Bromberger \inst{1} \and Chaahat Jain\inst{1,2} \and Christoph Weidenbach\inst{1}}
\authorrunning{Bromberger et al.}
%
\institute{Max Planck Institute for Informatics, Saarbr\"ucken, Germany\\
\email{\{mbromber, cjain, weidenbach\}@mpi-inf.mpg.de} \and
Graduate School of Computer Science, Saarbr\"ucken, Germany}

\maketitle

\begin{abstract}
  We show that SCL(FOL) can simulate the derivation of non-redundant clauses by superposition for first-order logic without equality.
  Superposition-based reasoning is performed with respect to a fixed reduction ordering. 
  The completeness proof of superposition relies on the grounding of the clause set. 
  It builds a ground partial model according to the fixed ordering, where minimal false ground instances of clauses then
  trigger non-redundant superposition inferences.
  We define a respective strategy for the SCL calculus such that clauses learned by SCL and superposition inferences coincide.
  From this perspective the SCL calculus can be viewed as a generalization of the superposition calculus.
\end{abstract}

\section{Introduction}

Superposition~\cite{BachmairGanzinger94b,BachmairGanzinger01handbook,NieuwenhuisRubio01handbook} is currently considered as the prime calculus for first-order logic reasoning
where all leading first-order theorem provers implement a variant thereof~\cite{Korovin08,SchulzEtAl19,KovacsVoronkov13,WeidenbachEtAlSpass2009}.
More recently, the family of SCL calculi (Clause Learning from Simple Models, or just Simple Clause Learning) \cite{FioriWeidenbach19,BrombergerFW21,BrombergerSW22,LeidingerWeidenbach22,BrombergerEtAl2023arxiv}
was introduced.
There are first experimental results \cite{BrombergerEtAl22} available, and
first steps towards an overall implementation \cite{BrombergerEtAl22IJCAR,BrombergerGLW22}.

The main differences between superposition and SCL for first-order logic without equality are: (i)~superposition assumes a fixed ordering on literals whereas the ordering in SCL is dynamic and evolves out of
the satisfiability of clauses, (ii)~superposition performs single superposition left and factoring inferences whereas SCL typically performs several
such inferences to derive a single learned clause, (iii)~the superposition model operator is not effective on the non-ground clause level whereas the SCL model assumption
is effective.
For first-order logic without equality superposition reduces to ordered resolution combined with the powerful superposition redundancy criterion.
Our simulation result cannot be one-to-one because an SCL learned clause is typically generated by several superposition inferences and
superposition factoring inferences are performed by SCL only in the context of resolution inferences. The simulation result considers the
ground case, where the superposition strategy used in the completeness proof only triggers non-redundant inferences~\cite{BachmairGanzinger94b}.
We call this strategy SUP-MO, Def.~\ref{def:supstrat}. Overall first-order completeness is then obtained by a lifting argument to the non-ground clause level.
We actually show that a superposition refutation of some ground clause set can be simulated by an SCL refutation on the same clause set, such
that they coincide on all superposition left (ordered resolution) inferences. For the superposition calculus we refer to \cite{BachmairGanzinger94b} and
for SCL to \cite{BrombergerEtAl2023arxiv} where all main properties of both calculi have meanwhile been verified
inside the Isabelle framework~\cite{WaldmannTRB20,DeshernaisIsa23,SchlichtkrullEtAlIsa18}.

For example, consider a superposition refutation of the simple ground clause set\newline
\centerline{$ N^0_{\text{SUP}} \{(C_1)~P(a)\lor P(a), \quad (C_2)~\neg P(a) \lor Q(b), \quad (C_3)~\neg Q(b)\}$}
with respect to a KBO~\cite{KnuthBendix70}, where all symbols have weight one, and precedence $a\prec b\prec P \prec Q$. Superposition
generates only non-redundant clauses. Then with respect
to the usual superposition ordering extension to literals and clauses we get $(C_1) \prec_{\text{KBO-SUP}} (C_2) \prec_{\text{KBO-SUP}} (C_3)$
and the superposition model operator produces the Herbrand model $N^0_{_{\text{SUP}},\cal I} = \emptyset$. Now clause $(C_1)$ is the minimal false clause,
triggering a factoring inference resulting in $(C_4)~P(a)$ and clause set $N^1_{\text{SUP}} = N^0_{\text{SUP}} \cup \{(C_4)~P(a)\}$. The clause $P(a)$ cannot be
derived by SCL because factoring is only preformed in the context of resolution inferences. Now $(C_4)$ is the smallest clause in $N^1_{\text{SUP}}$
and the superposition model operator produces $N^1_{_{\text{SUP}},\cal I} = \{P(a), Q(b)\}$ with minimal false clause $(C_3)$. A superposition left
inference between $(C_3)$ and $(C_2)$ generates $(C_5)~\neg P(a)$ and $N^2_{\text{SUP}} = N^1_{\text{SUP}} \cup \{(C_5)~\neg P(a)\}$. The generation of $\neg P(a)$ can now be simulated by SCL by constructing
the SCL trail $[P(a)^1 Q(b)^{\{\neg P(a) \lor Q(b)\}}]$ out of $N^0_{\text{SUP}} = N^0_{\text{SCL}}$ leading to the learned clause $(C_5)~\neg P(a)$
and respective clause set $N^2_{\text{SCL}} = N^0_{\text{SCL}} \cup \{(C_5)~\neg P(a)\}$.
Note that $P(a)$ could have also been propagated, see Sec.~\ref{sec:prelim} rule Propagate, but this would eventually not lead to the learned clause $(C_5)~\neg P(a)$ but $\bot$.
Finally, the superposition model operator produces $N^2_{_{\text{SUP}},\cal I} = \{P(a), Q(b)\}$ with minimal false clause $(C_5)$ and infers $\bot$.
The SCL simulation generates the trail $[P(a)^{\{P(a)\}}]$ and then learns $\bot$ as well out of a conflict with $(C_5)$. Note that this SCL trail is based
on a factoring of $(C_1)$ to $P(a)$ that was the explicit first step  of the superposition refutation.
Recall that by using an exhaustive propagation strategy, SCL would start with the trail  $[P(a)^{P(a)} Q(b)^{\{\neg P(a) \lor Q(b)\}}]$ and immediately derive $\bot$. Exhaustive propagation is
not a good strategy in general, because first-order logic clauses may enable infinitely many propagations. Therefore, the \emph{regular} strategy defined in \cite{BrombergerEtAl2023arxiv}
does not require exhaustive propagation, but guarantees non-redundant clause learning. The SCL-SUP strategy, Def.~\ref{def:sclsupstrat1}, and Def.~\ref{def:sclsupstrat2}, simulating superposition SUP-MO runs is also
a regular strategy, Lem.~\ref{lem:sclsupreasstrat}.

The paper is now organized as follows. After repetition of the needed concepts of SCL and superposition, Sec.~\ref{sec:prelim},
the simulation result is contained in Sec.~\ref{sec:sclsimulates}. We show that any superposition refutation of a ground clause set producing
only non-redundant inferences through the SUP-MO strategy, can be simulated via the SCL-SUP strategy.
Based on the 14 simulation invariants of Def.~\ref{def:sclsupstate}, we show the invariants by an inductive argument on the length of the superposition refutation,
starting from the initial state, Lem.~\ref{lem:sclsupinisimul},
for intermediate superposition inference steps Lem.~\ref{lem:sclsuppressimul} until the final refutation Lem.~\ref{lem:sclsupadvances}, and Lem.~\ref{lem:sclsupfinal}.
For the simulation we do not consider selection in superposition inferences in favor of a less complicated presentation. The paper ends with a discussion
of the obtained results.
All proofs have been moved to an appendix in favor of improved readability. 

\section{Preliminaries} \label{sec:prelim}

We assume a first-order language without equality where
$N$ denotes a clause set;
$C, D$ denote clauses;
$L, K, H$ denote literals;
$A, B$ denote atoms;
$P, Q, R$ denote predicates;
$t, s$ terms;
$f, g, h$ function symbols;
$a, b, c$ constants;
and $x, y, z$ variables.
Atoms, literals, clauses and clause sets are considered as usual, where
in particular clauses are identified both with their disjunction and multiset
of literals~\cite{BrombergerEtAl2023arxiv}.
The complement of a literal is denoted by the function $\comp$.
The function $\atom(L)$ denotes the atomic part of a literal.
Semantic entailment $\models$ is defined as usual where variables in clauses
are assumed to be universally quantified.
Substitutions $\sigma, \tau$ are total mappings from variables to terms, where
$\dom(\sigma) := \{x \mid x\sigma\neq x\}$ is finite and $\cdom(\sigma) := \{ t\mid x\sigma = t, x\in\dom(\sigma)\}$.
Their application is extended to literals, clauses, and sets of such objects in the usual way.
A term, atom, clause, or a set of these objects is \emph{ground} if it does not contain any variable.
A substitution $\sigma$ is \emph{ground} if $\cdom(\sigma)$ is ground. A substitution $\sigma$ is \emph{grounding}
for a term $t$, literal $L$, clause $C$ if $t\sigma$, $L\sigma$, $C\sigma$ is ground, respectively.
The function $\mgu$ denotes the \emph{most general unifier} of two terms, atoms, literals.
We assume that any $\mgu$ of two terms or literals does not introduce any fresh variables and is idempotent.
A \emph{closure} is denoted as $C\cdot\sigma$ and is a pair of a clause $C$ and a substitution $\sigma$ that is grounding for $C$.
The function $\gnd$ returns the set of all ground instances of a literal, clause, or clause set with respect
to the signature of the respective clause set.

A \emph{(partial) model} $M$ for a clause set $N$ is a satisfiable set of ground literals.  A ground clause $C$ is true in
$M$, denoted $M\models C$, if $C\cap M \not= \emptyset$, and false otherwise. A ground clause set $N$ is true in $M$, denoted $M\models N$ if
all clauses from $N$ are true in $M$.
A \emph{(partial) Herbrand model} $I$ for a clause set $N$ is a set of ground atoms.  A ground clause $C$ is true in
$I$, denoted $I \models_H C$, if there is an atom $A\in C$ such that $A\in I$, or there is a negative literal $\neg A\in C$ such that $A\not\in I$,
and false otherwise.
A ground clause set $N$ entails a ground clause $C$, denoted $N \models C$, if  $M\models C$ implies $M \models \{C\}$ for all models $M$.

We identify sets and sequences whenever appropriate. However, the trail of an SCL run is always a sequence of ground literals.

Let  $\prec$ denote a well-founded, total, strict ordering on
ground literals.
This ordering is then lifted to clauses and clause sets by its respective multiset extension. We
overload $\prec$ for literals, clauses, clause sets if the meaning is clear from the context.
The ordering is lifted to the non-ground case via instantiation: we define $C \prec D$
if for all grounding substitutions $\sigma$ it holds $C\sigma \prec D\sigma$.
We define $\preceq$ as the reflexive closure of $\prec$ and $N^{\preceq C} := \{D \mid D\in N \;\text{and}\; D\preceq C\}$.

\begin{definition}[Clause Redundancy] \label{prelim:def:redundancy}
  A ground clause $C$ is \emph{redundant} with respect to a ground clause
  set $N$ and an order $\prec$ if $N^{\preceq C} \models C$.
  A clause $C$ is \emph{redundant}  with respect to a clause set
  $N$ and an order $\prec$  if for all $C' \in \gnd(C)$ it holds that $C'$ is
  redundant with respect to $\gnd(N)$.
\end{definition}

Let $\prec_B$ denote a  well-founded, total, strict ordering on
ground atoms such that for any ground atom $A$ there are only
finitely many ground atoms $B$ with $B\prec_B A$.
For example, an instance of such an ordering could be KBO without zero-weight symbols.
(Note that LPO does not satisfy the last condition of a $\prec_B$ ordering although it is a well-founded, total, strict ordering.)
The ordering $\prec_B$ is lifted to literals
by comparing the respective atoms and if the atoms of two literals are the same, 
then the negative version of the literal is larger than the positive version. 
It is lifted to clauses by a multiset extension.

\paragraph{The SCL(FOL) Calculus:}
The inference rules of SCL(FOL)~\cite{BrombergerEtAl2023arxiv} are represented by
an abstract rewrite system.
They operate on a problem state, a six-tuple
$(\Gamma; N; U; \beta; k; D)$ where $\Gamma$ is a sequence
of annotated ground literals, the \emph{trail};
$N$ and $U$ are the sets of \emph{initial} and \emph{learned}
clauses; $\beta$ is a ground literal limiting the size of the trail; $k$ counts the number of decisions; and
$D$ is either $\top$, $\bot$ or a clause closure $C\cdot\sigma$ such that
$C\sigma$ is ground and false in $\Gamma$.
Literals in $\Gamma$ are either annotated with
a number, also called a level; i.e., they have the form $L^k$
meaning that $L$ is the $k$-th guessed decision
literal, or they are annotated with a closure that
propagated the literal to become true.
A ground literal $L$ is of
\emph{level} $i$ with respect to a problem state
$(\Gamma; N; U; \beta; k; D)$ if $L$ or $\comp(L)$ occurs
in $\Gamma$ and the first decision literal left from
$L$ ($\comp(L)$) in $\Gamma$, including $L$, is annotated with $i$.
If there is no such decision literal then its level
is zero. A ground clause $D$ is of \emph{level} $i$
with respect to a problem state $(\Gamma; N; U; \beta; k; D)$
if $i$ is the maximal level of a literal in $D$. The level of the empty clause $\bot$ is 0.
Recall $D$ is a non-empty closure or $\top$ or $\bot$.
Similarly, a trail $\Gamma$ is of level $i$ if the maximal literal in $\Gamma$ is of level $i$.

A literal/atom $L$/$A$ is \emph{undefined} in $\Gamma$
if neither $L$/$A$ nor $\comp(L)$/$\comp(A)$ occur in $\Gamma$.
The start state of SCL is $(\epsilon;N;\emptyset;\beta;0;\top)$ for some initial clause set $N$ and bound $\beta$.
The below rules are exactly the rules from \cite{BrombergerEtAl2023arxiv} and serve as a reference for our simulation proof in Sec.~\ref{sec:sclsimulates}.

\smallskip
\shortrules{Propagate}
{$(\Gamma;N;U;\beta;k;\top)$}
{$(\Gamma, L\sigma^{(C_0\lor L){\delta}\cdot\sigma};N;U;\beta;k;\top)$}
{provided $C\lor L\in (N\cup U)$, $C = C_0 \lor C_1$, $C_1\sigma = L\sigma \lor \dots \lor L\sigma$,
  $C_0\sigma$ does not contain $L\sigma$, {$\delta$ is the mgu of the literals in $C_1$ and $L$}, $(C\lor L)\sigma$ is ground, $(C\lor L)\sigma \prec_\beta \{\beta\}$,
  $C_0\sigma$ is false under $\Gamma$, and $L\sigma$ is undefined in $\Gamma$}{SCL}{12}

\smallskip
\shortrules{Decide}
{$(\Gamma;N;U;\beta;k;\top)$}
{$(\Gamma,L\sigma^{k+1};N;U;\beta;k+1;\top)$}
{provided $\atom(L)$ occurs $C$ for a $C \in (N\cup U)$, $L\sigma$ is a ground literal undefined in $\Gamma$, and  $L\sigma \prec_\beta \beta$}{SCL}{12}

\smallskip
\shortrules{Conflict}
{$(\Gamma;N;U;\beta;k;\top)$}
{$(\Gamma;N;U;\beta;k;D\cdot\sigma)$}
{provided $D\in (N\cup U)$, $D\sigma$ false in $\Gamma$
  for a grounding substitution $\sigma$}{SCL}{12}

\smallskip
\shortrules{Skip}
{$(\Gamma, L;N;U;\beta;k;D\cdot\sigma)$}
{$(\Gamma;N;U;\beta;k-i;D\cdot\sigma)$}
{provided $\comp(L)$ does not occur in $D\sigma$, if $L$ is a decision literal then $i=1$, otherwise $i=0$}{SCL}{11}

\smallskip
\shortrules{Factorize}
{$(\Gamma;N;U;\beta;k;(D\lor L \lor L')\cdot\sigma)$}
{$(\Gamma;N;U;\beta;k; (D\lor L)\eta\cdot\sigma)$}
{provided $L\sigma = L'\sigma$, $\eta=\mgu(L,L')$}{SCL}{11}

\smallskip
\shortrules{Resolve}
{$(\Gamma, L\delta^{(C\lor L)\cdot\delta};N;U;\beta;k;(D\lor L')\cdot\sigma)$ \\ \hspace*{2.46em} }
{$(\Gamma, L\delta^{(C\lor L)\cdot\delta};N;U;\beta;k;(D\lor C)\eta\cdot\sigma\delta)$}
{provided $L\delta = \comp(L'\sigma)$,
 $\eta=\mgu(L,\comp(L'))$}{SCL}{13}

\smallskip
\shortrules{Backtrack}
{$(\Gamma_0,K,\Gamma_1,\comp(L\sigma)^k;N;U;\beta;k;(D\lor L)\cdot\sigma)$ \\ \hspace*{2.8em}}
{$(\Gamma_0;N;U\cup\{D\lor L\};\beta;j;\top)$}
{provided $D\sigma$ is of level $i'<k$,
 and $\Gamma_0,K$ is the minimal trail subsequence such that there is
 a grounding substitution $\tau$ with $(D \lor L)\tau$ is false in $\Gamma_0,K$ but not in  $\Gamma_0$, and $\Gamma_0$ is of level $j$}{SCL}{13}

A sequence of rule applications of a particular calculus is called a \emph{run} of the calculus.
A \emph{strategy} for a calculus restricts the set of runs we actually allow by imposing further conditions 
on the allowed rule applications.

\begin{definition}[SCL Runs] \label{def:fol:sclrun}
  A sequence of SCL rule applications is called a \emph{reasonable run} if the rule Decide does not enable
  an immediate application of rule Conflict. A sequence of SCL rule applications is called
a \emph{regular run} if it is a reasonable run and the rule Conflict has precedence over all
other rules.
\end{definition}

All regular SCL runs are sound, 
only derive non-redundant clauses, always terminate, 
and SCL with a regular strategy is refutationally complete (for first-order logic without equality)~\cite{BrombergerEtAl2023arxiv}.

\paragraph{The Superposition Calculus:}
Superposition~\cite{BachmairGanzinger94b,BachmairGanzinger01handbook,NieuwenhuisRubio01handbook} 
is a calculus for first-order logic reasoning that also infers/learns new clauses like SCL. 
In contrast to SCL, it does these inferences based on a static ordering $\prec$ and, at the level of inference rules, independent of a partial model.
A permissible ordering $\prec$ for the superposition calculus is always a well-founded, total, strict ordering on ground literals.
This ordering is then lifted to clauses and clause sets by its respective multiset extension.
A problem state in the superposition calculus is just a set  $N$ of clauses.
The start state the initial clause set.
Due to the restriction to first-order logic without equality, 
the most basic version of the superposition calculus consists just of the following two rules (without selection):

\shortrules{Superposition Left}{$(N\disjun\{C_1\lor P(\tenum{t}{n}), C_2\lor \neg P(\tenum{s}{n})\})$}{$(N\cup\{C_1\lor P(\tenum{t}{n}), C_2\lor \neg P(\tenum{s}{n})\}\cup\{(C_1\lor C_2)\sigma\})$}{where 
(i)~$P(\tenum{t}{n})\sigma$ is strictly maximal in $(C_1\lor P(\tenum{t}{n}))\sigma$ \newline
(ii)~$\neg P(\tenum{s}{n})\sigma$ is maximal,
(iii)~$\sigma$ is the mgu of $P(\tenum{t}{n})$ and $P(\tenum{s}{n})$}{SUP}{25}

\smallskip
\shortrules{Factoring}{$(N\disjun\{C\lor P(\tenum{t}{n})\lor P(\tenum{s}{n})\})$}{$(N\cup\{C\lor P(\tenum{t}{n})\lor P(\tenum{s}{n})\}\cup\{(C\lor P(\tenum{t}{n}))\sigma\})$}{where (i)~ $P(\tenum{t}{n})\sigma$ is maximal in $(C\lor P(\tenum{t}{n})\lor P(\tenum{s}{n}))\sigma$  \newline
(ii)~$\sigma$ is the mgu of $P(\tenum{t}{n})$ and $P(\tenum{s}{n})$}{SUP}{25}

Let $\sfac(C)$ represent a clause obtained by exhaustively applying superposition Factoring on $C$. Recall, that superposition
Factoring only applies to maximal positive literals. Let $\sfac(N)$ represent the clause set $N$ after every clause has been
exhaustively factorized by Superposition Factorization.

Although the superposition calculus itself is independent of a partial model and may learn non-redundant clauses,
the completeness proof of superposition in~\cite{BachmairGanzinger94b} is based on a strategy that 
builds ground partial models according to the fixed ordering $\prec$, 
where minimal false ground instances of clauses then trigger non-redundant superposition inferences.
Note that the completeness proof relies on a grounding of the clause set that may lead to infinitely many clauses.
However, 
the strategy from the completeness proof can also be seen as a superposition strategy 
for an initial clause set, where all clauses are already ground.
On ground, finite clause sets, superposition restricted to the strategy 
only infers non-redundant clauses, always terminates, 
and is complete.
The partial model needed in each step of the strategy is constructed according to the following model operator:

\begin{definition}[Superposition Model Operator] \label{def:supmodelop}
Let $N$ be a set of ground clauses. Then $N_I$ is the Herbrand model according to the superposition model operator for clause set $N$
and it is constructed recursively over the partial Herbrand models $N_C$ for all $C \in N$:
  
\renewcommand{\arraystretch}{1.5}
$\begin{array}{ll}
    N_C = \bigcup_{D \prec C}\delta_D \qquad\qquad N_I = \bigcup_{C \in N} \delta_C\\
    \delta_D = \begin{cases}
    \{B\} & \text{if $D = D' \lor B, B$ strictly maximal,  $N_D \not\models_H D$}\\
    \emptyset & \text{otherwise}
    \end{cases}\\
\end{array}$

We say that a clause $C$ is \emph{productive} (wrt.\ the model construction of a clause set $N$) if $\delta_C \neq \emptyset$.
We say that a clause $C$ \emph{produces} an atom $B$ (wrt.\ the model construction of a clause set $N$) if $\delta_C = \{B\}$.
\end{definition}

After constructing the model $N_I$ for a clause set $N$,
the strategy selects the smallest clause in $N$ that is false in $N_I$.
The strategy then selects a fitting inference rule
based on the reason why the clause is false in $N_I$.
The newly inferred clause either changes the model in the next step or changes the smallest clause that is false.
This is the strategy used in the superposition completeness proof~\cite{BachmairGanzinger94b}.

\begin{definition}[Minimal False Clause] \label{def:supminfalseclause}
  The minimal false clause $C \in N$ is the smallest clause in $N$ according to $\prec$ such that 
  $N_C \cup \delta_C \not\models_H C$.
\end{definition}

\begin{definition}[Superposition Model-Operator Strategy: SUP-MO] \label{def:supstrat}
  The \emph{superposition model-operator strategy} is defined over the minimal false clause with regards to the current clause set $N$.
  The strategy can encounter the following cases:
  \begin{description}
    \item[(1)] $N$ has no minimal false clause. Then $N$ is satisfied by $N_I$ and we can stop the superposition run.
    \item[(2)] The minimal false clause in $N$ is $\bot$. Then $N$ is unsatisfiable, which means we can also stop the superposition run.
    \item[(3)] $C$ is the minimal false clause in $N$, and it has a maximal literal $L$ that is negative. 
        Then there must be a clause $D \in N$ with $D \prec C$, a strictly maximal literal $\comp(L)$, and $\delta_D = \{\comp(L)\}$.
        In this case, the strategy applies as its next step Superposition Left to $C$ and $D$.
    \item[(4)] $C$ is the minimal false clause in $N$, and it has a maximal literal $L$ that is positive. 
        Then $L$ is not strictly maximal in $C$ and the strategy applies Factoring to $C$.
  \end{description}
\end{definition}

The first two cases of the SUP-MO strategy also describe its final states according to~\cite{BachmairGanzinger94b}.
In all other states there is always exactly one rule applicable according to the SUP-MO strategy, 
which also means that SUP-MO is never stuck.

\begin{lemma}[SUP-MO Applicability] \label{lem:supstratapp}
  Let $N$ be a set of ground clauses.   
  If $N$ has a minimal false clause $C \neq \bot$, then there exists exactly one rule applicable to $N$ according to the SUP-MO strategy.
\end{lemma}

\section{SCL Simulates Superposition} \label{sec:sclsimulates}

In general, it is not possible to simulate all inferences of the superposition calculus 
with SCL because SCL only learns/infers non-redundant clauses, 
whereas syntactic superposition inferences have no such guarantees.
Moreover, the inferences by SCL are all based on conflicts according to a partial model driven by the satisfiability of clause instances, 
whereas the inferences by superposition are based on a static ordering $\prec$.
We can mitigate these differences by restricting superposition with the SUP-MO strategy
because SUP-MO has non-redundancy guarantees and 
it infers new clauses based on minimal false clauses with respect to a ground partial model.

Let $N^0$ be a set of ground clauses, totally ordered by a superposition reduction ordering $\prec$. 
Let $N^i$ (for $i > 0$) be the result of $i$ steps of the superposition calculus applied to $N^0$ according to the SUP-MO strategy, i.e.,
$N^0 \Rightarrow_{\text{SUP-MO}} N^1 \Rightarrow_{\text{SUP-MO}} \ldots \Rightarrow_{\text{SUP-MO}} N^i$.
Again, all $N^i$ are sets of ground clauses, totally ordered by a superposition reduction ordering $\prec$. 
The SCL strategy SCL-SUP that simulates superposition restricted to SUP-MO runs is defined inductively on the clause ordering $\prec$.
To guide and to prove the correctness of our simulation, 
we assign to each SCL state and every clause some additional information.
For this purpose, every SCL state is annotated with a triple $(i,C,\gamma)$,
where $i$ is an integer that states that the SCL state simulates the superposition state $N^i$,
$C$ is the last clause that was used as a decision aid by the strategy,
$\gamma$ is a function such that $\gamma(C) = \sfac(C)$ if $\sfac(C) \in N^i$ and $\gamma(C) = C$ otherwise,
the SCL state also simulates the model construction for $N^i$ upto $N^i_{C'} \cup \delta_{C'}$, where $C' = \gamma(C)$.
The annotated states are written $(\Gamma;N^0;U;\beta;k;E)_{(i,C,\gamma)}$.
The overall start state is then $(\epsilon;N^0;\emptyset;\beta;0;\top)_{(0,\bot,\gamma)}$, where
we assume $\beta$ large enough so $A \prec_\beta \beta$ for all $A \in \atom(N^0)$, $\bot \not\in N^0$, and $\gamma(C) = \sfac(C)$ if $\sfac(C) \in N^0$ and $\gamma(C) = C$ otherwise.
We will later see that the annotated integer is not relevant for the actual choice of SCL rules by the SCL-SUP strategy but only 
to prove that the strategy actually simulates superposition.
Moreover, we define a new ordering $\prec_{\gamma}$ based on our superposition ordering $\prec$ and function $\gamma$ such that 
$C \prec_{\gamma} D$ if $\gamma(C) \prec \gamma(D)$. 


\begin{definition}[State Simulation] \label{def:sclsupstate}
  Let $(\Gamma;N^0;U;\beta;k;E)_{(i,D,\gamma)}$ be an SCL state for the input clauses $N^0$.
  Let $L$ be the maximal literal in $D$ if $D \neq \bot$ and the minimal literal according to $\prec$ otherwise.
  Let $N^0 \Rightarrow_{\text{SUP-MO}} N^1 \Rightarrow_{\text{SUP-MO}} \ldots \Rightarrow_{\text{SUP-MO}} N^i$ be the superposition run following the SUP-MO strategy
  starting from the input clause set $N^0$.
  Let $D' = \gamma(D)$.
  Then we say that the SCL state $(\Gamma;N^0;U;\beta;k;E)_{(i,D,\gamma)}$ \emph{simulates} $N^i$ and the model construction upto $N^i_{D'} \cup \delta_{D'}$ if
  \begin{description}
    \item[(i)] $\atom(N^0) = \atom(N^i) = \atom(N^0 \cup U)$, $A \prec_\beta \beta$ for all $A \in \atom(N^0)$, and $D \in \{\bot\} \cup N^0 \cup U$
    \item[(ii)] $\sfac(N^0 \cup U) \subseteq \sfac(N^i)$ and $\gamma(C) \in N^i$ for all $C \in N^0 \cup U$ and $\gamma(C) = \sfac(C)$ or $\gamma(C) = C$.
    \item[(iii)] for all $C \in N^i$ there exists a $C' \in N^0 \cup U \cup \{E\}$ such that $\sfac(C) = \sfac(C')$ if the maximal literal in $C$ is positive
    \item[(iv)] for all $C \in N^i$ there exists a $C' \in N^0 \cup U \cup \{E\}$ such that $C' \models C$ and $\gamma(C') \preceq C$
    \item[(v)] for all atoms $A$ occurring in $N^0$: $A\in N_{D'}\cup\delta_{D'}$ iff $A\in\Gamma$
    \item[(vi)] for all atoms $A$: $\neg A \in \Gamma$ iff $A \prec L$ and $A\not\in N_{D'}$
    \item[(vii)] for every literal $L$ in $\Gamma$, i.e., $\Gamma = \Gamma', L, \Gamma''$, 
                and all literals $L'$ in $\Gamma'$, $\atom(L') \prec \atom(L)$
    \item[(viii)] for every atom (= positive literal) $B$ in $\Gamma$, i.e., $\Gamma = \Gamma', B, \Gamma''$, 
                 there exists $C \in N^0 \cup U$ and a $C' \in N^i$ such that $\gamma(C) = \sfac(C) = \sfac(C') = C'$, and $C'$ produces $B$, i.e., $\delta_{C'} = \{B\}$
    \item[(ix)] for every clause $C \in N^i$ with $C \preceq \gamma(D)$ that produces an atom $B$, i.e., $\delta_{C} = \{B\}$, 
                  there exists $C' \in N^0 \cup U$ such that $C = \gamma(C')$ and $C \preceq_{\gamma} D$.
    \item[(x)] $\Gamma$ contains only decisions if $E = \top$
    \item[(xi)] $E \not \in \{\top, \bot\}$ iff $\Gamma = \Gamma' B^{\sfac(D)}$, $\Gamma'$ contains only decisions, there exists $E' \in N^i$ where $\gamma(E) = E = E'$ is the minimal false clause in $N^i$, and $\neg B \in E$  
    \item[(xii)] $\Gamma \models C$ for all $C \in N^0 \cup U$ with $C \preceq_\gamma D$
    \item[(xiii)] Conflict is not applicable to $(\Gamma;N^0;U;\beta;k;E)_{(i,D,\gamma)}$.
    \item[(xiv)] $\bot \not\in N^0 \cup U$ and $E = \bot$ iff $\Gamma = \epsilon$ and $\bot \in N^i$
  \end{description}
\end{definition}

The above invariants can be summarized as follows:
(i)~All ground atoms encountered are known from the start and the trail bound $\beta$ is large enough so SCL can Decide/Propagate them.
(ii)-(iv)~Every initial clause $C$ or inferred clause by SUP-MO must coincide with an initial clause $C'$ or learned clause by SCL; 
this means on the one hand that for every clause $C$ learned by SCL-SUP, SUP-MO infers a clause $C'$ that is identical up to factoring;
on the other hand it means that for every clause $C$ inferred by SUP-MO, SCL-SUP learns a clause $C'$ that entails $C$ (i.e. $C' \models C$) and is at most as large as $C$ wrt.\ $\gamma$.
(v)-(ix)~The partial model constructed by SCL-SUP and SUP-MO coincide and any atom $B$ in $N_C \cup \delta_C$ produced by clause $D$ 
has a clause $D'$ on the SCL side that could propagate $B$ and vice versa.
(x)-(xiii)~Ensure that any Conflict in SCL-SUP corresponds to a minimal false clause and that the trail is always constructed in such a way 
that the Resolve applications per Conflict call are limited to the maximal literal in the conflict; 
this property is needed or the next clause that would be learned by SCL no longer coincides with the clauses learned by SUP-MO.
(xiv)~Describes the final state in case the input clause set is unsatisfiable.

Now that we have defined how an SCL state must look like in order to simulate a superposition state, 
we define SCL-SUP, the SCL strategy that eventually simulates a SUP-MO run.
First, note that not all states visited by SCL-SUP satisfy the the invariants of Def.~\ref{def:sclsupstate}.
However, the invariants hold again after each so-called \emph{atomic sequence} of SCL-SUP steps.
Second, one atomic sequence of SCL-SUP steps may skip over several successive superposition states.
The reason is that SCL can and must skip all steps of SUP-MO that occur because the maximal literal in a clause is not strictly maximal, i.e.,
superposition Factoring steps.
SCL performs factoring implicitly in its Propagation rule so SCL never has to explicitly simulate case (4) of Def.~\ref{def:supstrat}.
Third, definition of the SCL-SUP strategy is split in two parts and each part describes some atomic sequences of SCL-SUP steps.

\begin{definition}[SCL Superposition Strategy: SCL-SUP Part 1] \label{def:sclsupstrat1}
  Let $S_0 = (\Gamma;N^0;U;\beta;k;\top)_{(i,C,\gamma)}$ be an SCL state with additional annotations for the strategy.
  Let $D$ be the next largest clause from $C$ in the ordering $\prec_{\gamma}$ with respect to the ground clause set $N^0 \cup U$. 
  Let $L$ be the maximal literal of $D$.
  Let $[\lnot A_1, \lnot A_2,\dots \lnot A_n]$ be all negative literals such that for all $i$ we have $A_i \prec L$, all $A_i$ undefined in $\Gamma$, $A_i$ occurs in $N^0 \cup U$, and  $A_i\prec A_{i+1}$.
  Let $D' = \gamma(D)$ be in $N^i$ such that $\sfac(D) = \sfac(D')$.
  Let $j_0 + 1$ be the number of occurrences of $L$ in $D'$ and $j = i + j_0$.
  Then the \emph{SCL Superposition Strategy} (SCL-SUP) performs the following steps to $S_0$ (possibly without any actual SCL rule applications, just changing the state annotation):
  \begin{description}
    \item[(1)] First decide all literals $[\lnot A_1, \lnot A_2,\dots \lnot A_n]$ in order, i.e., 
              $S_0  \Rightarrow^{*\;\text{Decide}}_{\text{SCL-SUP}} S_1$, 
              where $S_1 = (\Gamma,\neg A^{k+1}_1,\ldots, \neg A_n^{k+n};N^0;U;\beta;k+n;\top)_{(i,D,\gamma)}$.
    \item[(2a)] If the maximal literal $L$ in $D$ is positive (i.e., $L = B$), \newline
                $\Gamma,\neg A^{k+1}_1,\ldots, \neg A_n^{k+n} \not\models D$, 
                and Conflict is not applicable to\newline 
                $S_2 = (\Gamma,\neg A^{k+1}_1,\ldots, \neg A_n^{k+n}, B^{k+n+1};N^0;U;\beta;k+n+1;\top)_{(j,D,\gamma')}$, 
                then decide $B$, i.e, $S_1  \Rightarrow^{\text{Decide}}_{\text{SCL-SUP}} S_2$, where 
                $\gamma'$ is the same as $\gamma$ except that $\gamma'(D) = \sfac(D)$.
    \item[(2b)] If the maximal literal $L$ in $D$ is positive (i.e., $L = B$), \newline
                $\Gamma,\neg A^{k+1}_1,\ldots, \neg A_n^{k+n} \not\models D$, 
                and $E$ is the smallest clause in $N^0 \cup U$ that is false in wrt.\ $\Gamma,\neg A^{k+1}_1,\ldots, \neg A_n^{k+n},B^{\sfac(D)}$,
                then propagate $B$ and apply Conflict to $E$, i.e, 
                $S_1  \Rightarrow^{\text{Propagate}}_{\text{SCL-SUP}} S'_2  \Rightarrow^{\text{Conflict}}_{\text{SCL-SUP}} S_2$, \newline 
                where $S_2 = (\Gamma,\neg A^{k+1}_1,\ldots, \neg A_n^{k+n},B^{\sfac(D)};N^0;U;\beta;k+n;E)_{(j,D,\gamma')}$ and 
                $\gamma'$ is the same as $\gamma$ except that $\gamma'(D) = \sfac(D)$.
    \item[(2c)] Otherwise, $S_2 = S_1$ and no further rules have to be applied.
  \end{description}
  A (potentially empty) sequence of SCL rule applications according to SCL-SUP 
  is called an \emph{atomic sequence} of SCL-SUP steps if it starts from a state $S_0$ and ends in a state $S_2$ outlined in the cases (2a-c).
\end{definition}

The first part of the strategy simulates the recursive construction of the partial model used in the SUP-MO strategy (see Def.~\ref{def:supmodelop}).
It assumes that the model is already constructed up to the current annotated clause $C$ and 
extends this model for the next largest clause $D \in (N^0 \cup U)$.
To this end, it uses the rule Decide in step (1) to set all atoms $A$ to false that are still undefined but can no longer be produced by any clause greater or equal to $D$.
Next the strategy makes a case distinction.
Step (2a) handles the case where $D$ corresponds to a clause $D'$ in the superposition state (modulo some Factoring steps skipped by SCL) that produces atom $B$; 
SCL-SUP then adds $B$ to the trail with the rule Decide because producing/adding this atom does not falsify a clause.
Step (2b) handles a similar case compared to step (2a); 
but in this case producing/adding the atom $B$ to the trail results in a minimal false clause $E$;
in order to force a resolution step between clause $D$ and $E$, SCL-SUP first uses Propagate to add $B$ to the trail and then applies conflict to $E$.
Step (2c) handles the case where $D$ corresponds to a clause $D'$ that will not produce an atom $B$ even modulo some Factoring steps; 
in this case no further SCL rule applications are necessary as the SUP-MO model will not change.
Note that the annotated function $\gamma$ is needed so the SCL state knows when the superposition state would have applied Factoring to a clause $C$, 
which also means that it is now treated as its factorized version $\gamma(C)=\sfac(C)$ in our inductive clause ordering.

\begin{example}\label{ex:strat1}
Let us now further demonstrate the three different cases of the first part of the SCL-SUP strategy with the help of an example.
Let $N^0$ be our initial set of clauses:\newline
\centerline{$ N^0 = \{ (C_1)~P(a), \quad (C_2)~\neg P(b) \lor Q(a), \quad (C_3)~\neg P(a) \lor Q(a) \lor Q(a),$}\newline
\centerline{$ (C_4)~P(a) \lor \neg Q(a), \quad (C_5)~\neg P(a) \lor \neg Q(a)\}$}
We compare the run of SCL-SUP for $N^0$ with the run of SUP-MO for $N^0$ to demonstrate that both runs coincide.
As superposition ordering, we choose an LPO with precedence $a \prec b \prec P \prec Q$.
This means that the atoms are ordered $P(a) \prec P(b) \prec Q(a) \prec Q(b)$ and 
the clauses in $N^0$ are ordered $C_1 \prec C_2 \prec C_3 \prec C_4 \prec C_5$.
The initial SUP-MO state is simply the clause set $N^0$ and the 
initial SCL-SUP state is $(\epsilon,N^0,\emptyset,\beta,0,\top)_{(0,\bot,\gamma_0)}$, where $\gamma_0(C) = C$ for all clauses $C$.
In the first step of SCL-SUP, SCL-SUP first selects the clause $C_1$ as its new decision aid because it is the next largest clause in $N^0$ compared to $\bot$.
Then SCL-SUP continues with step (1) of Def.~\ref{def:supmodelop}.
In this step SCL-SUP does nothing because there are no atoms smaller than $P(a)$.
Next, SCL-SUP detects that the maximal literal of $C_1$ is positive, $\epsilon \not\models C_1$, and that the trail $[P(a)^1]$ does not result in a conflict.
Therefore, SCL-SUP follows step (2a) of Def.~\ref{def:supmodelop} and Decides $P(a)$, which results in the state 
$([P(a)^1],N^0,\emptyset,\beta,1,\top)_{(0,C_1,\gamma_0)}$.
Meanwhile, SUP-MO starts with constructing a model for $N^0$ starting with the clause $C_1$.
The result is that $C_1$ is productive and $\delta_{C_1} = \{P(a)\}$ and $N^0_{C_1} = \emptyset$, which coincides with our new SCL trail.

SCL-SUP considers the clause $C_2$ as its new decision aid and continues with step (1) of Def.~\ref{def:supmodelop}.
This time there is an atom smaller than the maximal literal of $C_2$ namely $P(b)$.
Therefore, SCL-SUP Decides $\neg P(b)$ in step (1) of Def.~\ref{def:supmodelop}, which results in 
$([P(a)^1,\neg P(b)^2],N^0,\emptyset,\beta,2,\top)_{(0,C_2,\gamma_0)}$.
Next, SCL-SUP detects that the maximal literal of $C_2$ is positive but that $[P(a)^1, \neg P(b)^2] \models C_2$.
Therefore, SCL-SUP follows step (2c) of Def.~\ref{def:supmodelop} and ends this atomic sequence immediately.
SUP-MO continues the model construction for $N^0$ with the clause $C_2$.
The clause $C_2$ is not productive because $N^0_{C_2} \models_H C_2$, where $N^0_{C_2} = \delta_{C_1} = \{P(a)\}$ and $\delta_{C_2} = \emptyset$, 
which again coincides with our new SCL trail as Herbrand models do not explicitly define atoms assigned to false.

SCL-SUP now considers the clause $C_3$ as its new decision aid and continues with step (1) of Def.~\ref{def:supmodelop}.
In this step SCL-SUP does nothing because all atoms smaller than $Q(a)$ are already assigned.
Next, SCL-SUP detects that the maximal literal of $C_3$ is positive, $[P(a)^1,\neg P(b)^2] \not\models C_3$, and that the clause $C_5$ is false with respect to the trail $[P(a)^1,\neg P(b)^2,Q(a)^{\sfac(C_3)}]$.
Therefore, SCL-SUP follows step (2b) of Def.~\ref{def:supmodelop}, i.e. it Propagates $P(a)$ and applies Conflict to $C_5$,
resulting in $([P(a)^1,\neg P(b)^2,Q(a)^{\sfac(C_3)}],N^0,\emptyset,\beta,2,C_3)_{(1,C_2,\gamma_1)}$,
where $\gamma_1$ is identical to $\gamma_0$ except that $\gamma_1(C_3) = \sfac(C_3) = \neg P(a) \lor Q(a)$.
Note that SCL-SUP must change the state annotations because the maximal literal in $C_3$ is not strictly maximal, 
so SCL-SUP skips and eventually silently performs the Factorization step performed by SUP-MO.
Note also that in the changed clause ordering $\prec_{\gamma_1}$ the order of $C_2$ and $C_3$ changed, i.e., $C_3 \prec_{\gamma_1} C_2$,
which corresponds to $\sfac(C_3) \prec C_2$.
Meanwhile, SUP-MO continues the model construction for $N^0$ with the clause $C_3$.
The clause $C_3$ is not productive because the maximal literal is not strictly maximal so $\delta_{(3)} = \emptyset$ and 
$N^0_{C_3} \cup \delta_{C_3} \not\models_H C_3$ so $C_3$ is the minimal false clause in $N^0$.
SUP-MO resolves this conflict by applying Factoring to $C_3$, 
which means SUP-MO infers the clause $C_6 = \sfac(C_3) = \neg P(a) \lor Q(a)$.
The new clause order in superposition state $N^1 = N^0 \cup \{C_6\}$ is $C_1 \prec C_6 \prec C_2 \prec C_3 \prec C_4 \prec C_5$, 
which matches the changed ordering $C_3 \prec_{\gamma_1} C_2$ because $C_6 = \gamma_1(C_3)$.
Next, SUP-MO updates its model construction for $N^1$. 
The result is that $C_1$ and $C_6$ are productive and that $N^1_{C_6} \cup \delta_{C_6} = \{P(a),Q(a)\}$, which matches the current SCL trail.
Moreover, if we continue the model construction upto $C_5$ then no new literals are produced and $C_5$ also turns into the minimal false clause for $N^1$.
\end{example}

\begin{definition}[SCL Superposition Strategy: SCL-SUP Part 2] \label{def:sclsupstrat2}
  Let $S_0 = (\Gamma, B^{\sfac(C)};N^0;U;\beta;k;E)_{(i,C,\gamma)}$ be an SCL state with $E \not\in \{\top,\bot\}$ and additional annotations for the strategy.
  Let $L = \neg B$ be the maximal literal of $E$.
  Let $\Gamma$ contain only decision literals.
  Let all atoms $A$ occurring in $N^0 \cup U$ with $A \prec B$ be defined in $\Gamma$ following the order $\prec$, i.e.,
  for all $A$ occurring in $N^0 \cup U$ with $A \prec B$ there exist $\Gamma'$ and $\Gamma''$ such that $\Gamma' = \Gamma', L_A, \Gamma''$, 
  $L_A = A$ or $L_A = \neg A$ and all atoms $A' \in N^0 \cup U$ with $A' \prec A$ are defined in $\Gamma'$.
  Let $E$ be contained in $N^i$.
  Let $j_0$ be the number of occurrences of $L$ in $E$ and $j = i + j_0$.  
  Let $\sfac(C) = C_1 \lor B$ and $E = E' \lor E''$, where $E''$ contains all occurrences of $L$ in $E$.
  Then the \emph{SCL Superposition Strategy} (SCL-SUP) performs the following steps to $S_0$:
  \begin{description}
    \item[(1)] First apply Resolve to $E$ until all occurrences of $L$ are resolved away, i.e., 
               $S_0  \Rightarrow^{*\;\text{Resolve}}_{\text{SCL-SUP}} S_1$, 
               where $S_2 = (\Gamma, B^{\sfac(C)};N^0;U;\beta;k;E_2)_{(j,C,\gamma)}$ and $E_2 = E' \lor C_1 \lor \ldots \lor C_1$.
    \item[(2a)] If $E_2 = \bot$, then we apply Skip until the trail is empty and then stop the SCL run, i.e., 
               $S_2 \Rightarrow^{*\;\text{Skip}}_{\text{SCL-SUP}} S_5$, 
               where $S_5 = (\epsilon;N^0;U;\beta;0;\bot)_{(j,\bot,\gamma)}$.
    \item[(2b)] If $E_2 \neq \bot$, then $E_2$ has a maximal literal $L_1$. 
                Next the strategy applies Skip until $\comp(L_1)$ is the topmost literal on the trail, i.e., 
               $S_2  \Rightarrow^{*\;\text{Skip}}_{\text{SCL-SUP}} S_3$, 
               where $S_3 = (\Gamma_0,L_1^{k_1};N^0;U;\beta;k_1;E_2)_{(j,C,\gamma)}$.
               (Note that this step skips at least over the literal $B^{\sfac(C)}$).
    \item[(3)] Next apply Backtrack to $S_3$, i.e., $S_3  \Rightarrow^{\text{Backtrack}}_{\text{SCL-SUP}} S_4$,
               where $S_4 = (\Gamma_0;N^0;U \cup \{E_2\};\beta;k_1-1;\top)_{(j,C,\gamma)}$.
    \item[(4a)] If $L_1$ is a negative literal, continue with the following rule applications.
                Let $D$ be the smallest clause in $N^0 \cup U$ with maximum literal $\comp(L_1) = B_1$ 
                and $\Gamma_0 \not\models D$.
                Then Propagate $B_1$ from $D$, and apply Conflict to $E_2$, i.e.,
                $S_4 \Rightarrow^{\text{Propagate}}_{\text{SCL-SUP}} S''_4  \Rightarrow^{*\;\text{Conflict}}_{\text{SCL-SUP}} S_5$,
                where $S_5 = (\Gamma_0,B_1^{\sfac(D)};N^0;U \cup \{E_2\};\beta;k_1 - 1;E_2)_{(j,D,\gamma)}$.
    \item[(4b)] If $L_1$ is a positive literal (i.e., $L_1 = B$) and Conflict is not applicable to\newline 
                $S_5 = (\Gamma_0,B_1^{k_1};N^0;U \cup \{E_2\};\beta;k_1;\top)_{(j_2,E_2,\gamma')}$, 
                then decide $B$, i.e, $S_4  \Rightarrow^{\text{Decide}}_{\text{SCL-SUP}} S_5$, 
                where $j_1 + 1$ is the number of occurrences of $B_1$ in $E_2$, $j_2 = j + j_1$, and
                $\gamma'$ is the same as $\gamma$ except that $\gamma'(E_2) = \sfac(E_2)$.
    \item[(4c)] If $L_1$ is a positive literal (i.e., $L_1 = B$)
                and $E_3$ is the smallest clause in $N^0 \cup U$ that is false in 
                $S'_5 = (\Gamma_0,B_1^{\sfac(E_2)};N^0;U;\beta;k_1-1;\top)_{(j,E_2)}$,
                then propagate $B_1$ and apply Conflict to $E_3$, i.e, 
                $S_4 \Rightarrow^{\text{Propagate}}_{\text{SCL-SUP}} S'_5  \Rightarrow^{\text{Conflict}}_{\text{SCL-SUP}} S_5$, 
                where $S_5 = (\Gamma_0,B_1^{\sfac(E_2)};N^0;U;\beta;k_1-1;E_3)_{(j_2,E_2,\gamma')}$,
                $j_1 + 1$ is the number of occurrences of $B_1$ in $E_2$, $j_2 = j + j_1$, and 
                $\gamma'$ is the same as $\gamma$ except that $\gamma'(E_2) = \sfac(E_2)$.
  \end{description}
  A (potentially empty) sequence of SCL rule applications according to SCL-SUP 
  is called an \emph{atomic sequence} of SCL-SUP steps if it starts from a state $S_0$ and ends in a state $S_5$ outlined in the cases (2a) and (5a-c).
\end{definition}

The second part of the strategy simulates the actual inferences resulting from a minimal false clause found in step (2b) of Def.~\ref{def:sclsupstrat1} or 
found in steps (4a) and (4c) of Def.~\ref{def:sclsupstrat2}.
These inferences always correspond to Superposition Left steps of the SUP-MO strategy that 
resolve minimal false clauses $E'$ in $N^i$ with maximal literal $\neg B$
with the clause $C'$ in $N^i$ that produced $B$.
Note however that SCL-SUP may combine several Superposition Left steps of the SUP-MO strategy into one new learned clause.
This is the case whenever the maximal literal $\neg B$ in the minimal false clause $E'$ in $N^i$ is not strictly maximal.
In this case, the next minimal false clause $E''$ will always correspond to the last inferred clause,
the maximal literal of this clause will still be $\neg B$,
the clause producing $B$ will be again $C'$,
and therefore the next Superposition Left partner of $E''$ is also again $C'$.
Moreover, all of the skipped inferences are actually redundant with respect to the final inference $E'_2$ in this chain,
which explains why SCL-SUP is still capable of simulating SUP-MO although it skips the intermediate inferences.
The actual SCL-SUP clause $E_2$ corresponding to final SUP-MO inference $E'_2$ is computed in the steps (1) and (2) 
of Def.~\ref{def:sclsupstrat2} with greedy applications of the rules Resolve and Factorize.
The following steps of Def.~\ref{def:sclsupstrat2} take care of the four different cases how $E'_2$ changes the model and minimal false clause in $N^j$.
The first case is that $E'_2 = \bot$ so SUP-MO has reached a final state.
This case is handled by step (2a) of Def.~\ref{def:sclsupstrat2} that simply empties the trail with applications of the rule Skip 
so the resulting SCL state has the form of a SCL-SUP final state.
The second case is that the maximal literal $L_1$ in $E'_2$ is negative.
In this case, the model for $N^i$ and $N^j$ is still the same and just the minimal false clause changes to $E'_2$.
This case is handled by steps (2b)-(4a) of Def.~\ref{def:sclsupstrat2} that Backtrack before $\comp(L_1)$ was decided, 
propagate it instead and apply Conflict to $E_2$.
In the third and fourth case the maximal literal $L_1$ in $E'_2$ is positive.
In this case, the model for $N^i$ and $N^j$ actually changes because $E'_2$ is always productive.
Case (2b)-(4b) of Def.~\ref{def:sclsupstrat2} handles the case where producing $L_1$ leads to no new minimal false clause, 
and case (2b)-(4c) of Def.~\ref{def:sclsupstrat2} handles the case where it does.
Both cases work symmetrically to steps (2a) and (2b) of Def.~\ref{def:sclsupstrat1}.

\begin{example}
We continue Ex.~\ref{ex:strat1} to demonstrate cases (1)$\rightarrow$(4a) and (1)$\rightarrow$(2a) of the second part of the SCL-SUP strategy.
We left the runs in the SCL state $([P(a)^1,\neg P(b)^2,Q(a)^{\sfac(C_3)}],N^0,\emptyset,\beta,2,C_3)_{(1,C_2,\gamma_1)}$
that simulates the superposition state $N^1$, where\newline
\centerline{$ N^1 = \{ (C_1) P(a), \quad (C_2)~\neg P(b) \lor Q(a), \quad (C_3)~\neg P(a) \lor Q(a) \lor Q(a),$}\newline
\centerline{$ (C_4)~P(a) \lor \neg Q(a), \quad (C_5)~\neg P(a) \lor \neg Q(a),  \quad (C_6)~\neg P(a) \lor Q(a)\}$}
and $C_5$ became the minimal false clause in $N^1$ after $C_1$ and $C_6$ produced together the partial model $\{P(a),Q(a)\}$.
SUP-MO continues from the state $N^1$ by applying Superposition Left to $C_5$ and $C_6$.
In the new state $N^2 = N^1 \cup \{(C_7)~\neg P(a) \lor \neg P(a)\}$ the new clause order is 
$C_1 \prec C_7 \prec C_6 \prec C_2 \prec C_3 \prec C_4 \prec C_5$ 
and after constructing the model for $C_1$, which produces again $P(a)$, the clause $C_7$ becomes again the minimal false clause.
SCL-SUP follows (1) of Def.~\ref{def:sclsupstrat2} and applies Resolve to $C_5$ and $\sfac(C_3) = C_6$, resulting 
in the state $([P(a)^1,\neg P(b)^2,Q(a)^{\sfac(C_3)}],N^0,\emptyset,\beta,2,C_7)_{(2,C_2,\gamma_1)}$.
Then SCL-SUP continues with steps (2b) and (3) by applying Skip twice and Backtrack once to jump to the 
state $(\epsilon,N^0,\{C_7\},\beta,0,\top)_{(2,C_2,\gamma_1)}$.
Next, SCL-SUP continues with step (4a) because the maximal literal of $C_7$ is $\neg P(a)$ and therefore negative.
This means SCL-SUP will add $P(a)$ again to the trail but this time by applying Propagate to $C_1$ and afterwards it applies Conflict to $C_7$.
The resulting state $([P(a)^{\sfac(C_1)}],N^0,\{C_7\},\beta,0,C_7)_{(2,C_1,\gamma_1)}$ matches again the SUP-MO state $N^2$.

SUP-MO continues from the state $N^2$ by applying Superposition Left to $C_7$ and $C_1$, resulting in $N^3 = N^2 \cup \{(C_8) \neg P(a)\}$.
Since $C_8$ has the same maximal literal as $C_7$ it becomes automatically the next minimal false clause in $N^3$.
As a result, SUP-MO applies Superposition Left to $C_8$ and $C_1$, which returns $N^5 = N^3 \cup \{(C_9)~\bot\}$ a final state that proves the unsatisfiability of $N^0$.
Meanwhile, SCL-SUP simulates both Superposition Left steps with one atomic SCL-SUP sequence.
It starts with step (1) of Def.~\ref{def:sclsupstrat2} and applies Resolve twice, resulting 
in the state $([P(a)^1,\neg P(b)^2,Q(a)^{\sfac(C_3)}],N^0,\emptyset,\beta,2,\bot)_{(4,C_2,\gamma_1)}$.
Then it continues with step (2a) of Def.~\ref{def:sclsupstrat2} and applies Skip until the trail is empty.
The resulting state $(\epsilon,N^0,\emptyset,\beta,2,\bot)_{(4,\bot,\gamma_1)}$ is a final state and proves unsatisfiability of $N^0$.
\end{example}

\begin{example}
The next example demonstrates the atomic sequence (1)$\rightarrow$(4b) of the second part of the SCL-SUP strategy.
Let $N^0$ be our initial set of clauses:\newline
\centerline{$ N^0 = \{ (C_1) P(a), \quad (C_2)~\neg P(b), \quad (C_3)~\neg P(a) \lor Q(a), \quad (C_4)~P(b) \lor \neg Q(a)$}\newline
As superposition ordering, we choose an LPO with precedence $a \prec b \prec P \prec Q$.
This means that the atoms are ordered $P(a) \prec P(b) \prec Q(a) \prec Q(b)$ and 
the clauses in $N^0$ are ordered $C_1 \prec C_2 \prec C_3 \prec C_4$.
In order to keep the example short, we skip the initial SCL-SUP steps and 
continue directly with the state 
$S = ([P(a)^1,\neg P(b)^2,Q(a)^{\sfac(C_3)}],N^0,\emptyset,\beta,2,C_4)_{(0,C_3,\gamma)}$, where $\gamma(C) = C$ for all clauses $C$ and $\beta = Q(b)$.
This state simulates the superposition state $N^0$ upto the model construction for $C_3$, where $N^0_{C_3} \cup \delta_{C_3} = \delta_{C_1} \cup \delta_{C_3} = \{P(a), Q(a)\}$ 
and $C_4$ is the minimal false clause.
SUP-MO continues from the state $N^0$ by applying Superposition Left to $C_4$ and $C_3$.
In the new state $N^1 = N^0 \cup \{(C_5)~\neg P(a) \lor P(b)\}$ the new clause order is 
$C_1 \prec C_5 \prec C_2 \prec C_3 \prec C_4$ 
and the partial model upto $C_5$ is $N^0_{C_5} \cup \delta_{C_5} = \delta_{C_1} \cup \delta_{C_5} = \{P(a)\} \cup \{P(b)\}$,
which turns $C_2$ into the next minimal false clause.
SCL-SUP simulates the above steps by following the atomic sequence (1)$\rightarrow$(4b) of Def.~\ref{def:sclsupstrat2}.
The result is the state $([P(a)^1,P(b)^{\sfac(C_5)}],N^0,\{C_5\},\beta,1,C_2)_{(1,C_5,\gamma)}$ matching again our current superposition state and model.

Without clause $C_2$, SCL-SUP would apply the atomic sequence (1)$\rightarrow$(4a) of Def.~\ref{def:sclsupstrat2} to $S$, 
resulting in the state $([P(a)^1,P(b)^2],N^0 \setminus \{C_2\},\{C_5\},\beta,2,\top)_{(1,C_5,\gamma)}$.
This matches the state $N^1 \setminus \{C_2\}$ and its partial model upto $C_5$ that is still the same as for $N^1$
with the exception that it does not lead to a minimal false clause.
\end{example}

In order to actually show that every SCL-SUP run simulates a SUP-MO run, we need to prove three properties.
The first property is that each state visited by an SCL-SUP run must simulate a state visited by the corresponding SUP-MO run. 
Note that this property does not yet say anything about the order in which SCL-SUP simulates the SUP-MO states.
This property can also be seen as a soundness argument for our strategy.

\begin{lemma}[Initial SCL State Simulates Initial Superposition State] \label{lem:sclsupinisimul}
The initial SCL state $(\epsilon;N^0;\emptyset;\beta;0;\top)_{(0,\bot,\gamma)}$ simulates the initial superposition state $N^0$ and the model construction upto $N^0_{\bot} \cup \delta_{\bot}$
\end{lemma}

\begin{lemma}[SCL-SUP Preserves Simulation] \label{lem:sclsuppressimul}
Let the SCL state $S = (\Gamma;N^0;U;\beta;k;E)_{(i,C,\gamma)}$ simulate the superposition state $N^i$ and the corresponding model construction upto $N^i_{C'} \cup \delta_{C'}$,
where $C' = \gamma(C)$.
Let the SCL state $S' =(\Gamma';N^0;U';\beta;k';E')_{(j,D,\gamma')}$ be the result of one atomic sequence of SCL-SUP steps.
Then there exists a clause $D' \in N^j$ with $\gamma'(D) = D'$ 
and $S'$ simulates the superposition state $N^j$ and the model construction upto $N^j_{D'} \cup \delta_{D'}$.
\end{lemma}

The second property is that each atomic sequence of SCL-SUP steps always makes progress in the simulation. 
This means that each atomic sequence of SCL-SUP steps either advances the superposition state $N^i$ 
simulated by the current SCL state $S = (\Gamma;N^0;U;\beta;k;E)_{(i,D,\gamma)}$, i.e., it increases the annotated $i$,
or it still simulates the same superposition state $N^i$ but advances the simulation of the model construction operator, 
i.e. it increases the annotated clause $C$ and keeps $i$ the same.
Note that it can actually happen that an atomic sequence of SCL-SUP steps skips over several superposition states.
This property can also be seen as a termination argument for our strategy because SUP-MO always terminates on ground clause sets.

\begin{lemma}[SCL-SUP Advances the Simulation] \label{lem:sclsupadvances}
Let the SCL state $S = (\Gamma;N^0;U;\beta;k;E)_{(i,D,\gamma)}$ simulate the superposition state $N^i$ and the model construction upto $N^i_{D} \cup \delta_{D}$.
Let the SCL state $S' =(\Gamma';N^0;U';\beta;k';E')_{(j,D',\gamma')}$ be the next state reachable by one atomic sequence of SCL-SUP steps.
Then either $i < j$ or $i = j$ and $\gamma' = \gamma$ and $D \prec_{\gamma} D'$.
\end{lemma}

The last missing property shows that the SCL-SUP strategy can always advance the current SCL state 
whenever the simulated superposition state can be advanced by the SUP-MO strategy.
This means SCL-SUP is never stuck when SUP-MO can still progress.
These properties hold because the simulation invariants in Def.~\ref{def:sclsupstate} either correspond to a correct final state 
or they satisfy the preconditions of Def.~\ref{def:sclsupstrat1} or Def.~\ref{def:sclsupstrat2}.
This property can also be seen as a partial correctness argument for our strategy.

\begin{lemma}[SCL-SUP Correctness of Final States] \label{lem:sclsupfinal}
Let the SCL state $S = (\Gamma;N^0;U;\beta;k;E)_{(i,D,\gamma)}$ simulate the superposition state $N^i$ and the model construction upto $N^i_{\gamma(D)} \cup \delta_{\gamma(D)}$.
Let there be no more states reachable from $S$ following an atomic sequence of SCL-SUP steps.
Then $S$ is a \emph{final state}, i.e., either (i)~$E = \bot$, $D = \bot$, $\bot \in N^i$, and $N^0$ is unsatisfiable or 
(ii)~$\Gamma \models N^0$.
\end{lemma}

We can also show that any SCL-SUP run is also a regular run.
Although this is not strictly necessary for the simulation proof, 
it is beneficial because it means that SCL-SUP inherits many properties that hold for SCL restricted to a regular strategy.
For instance, that all learned clauses are non-redundant and that SCL-SUP always terminates.

\begin{lemma}[SCL-SUP is a Regular SCL Strategy] \label{lem:sclsupreasstrat}
SCL-SUP is a regular SCL strategy if it is executed on a state 
$S = (\Gamma;N^0;U;\beta;k;E)_{(i,C,\gamma)}$ that simulates a superposition state $N^i$ and the corresponding model construction upto $N^i_{\gamma(C)} \cup \delta_{\gamma(C)}$.
\end{lemma}

\section{Conclusion}

We have shown that the SCL(FOL) calculus~\cite{BrombergerEtAl2023arxiv} can simulate model driven superposition~\cite{BachmairGanzinger94b} refutations deriving only non-redundant
clauses. The superposition calculus cannot simulate SCL refutations due to its static a priori ordering.
In general, an SCL(FOL) learned clause is generated out of several resolution and factorization steps.
From this perspective the SCL(FOL) calculus is more general and flexible than the superposition calculus. Furthermore,
it only generates non-redundant clauses whereas any superposition implementation generates redundant clauses due
to the syntactic application of the superposition inference rules.

Selection in superposition can also be simulated, but requires an additional branch in the SCL-SUP strategy, because
selection of non-maximal, negative literals by superposition requires a different trail ordering for SCL in order to simulate a
respective superposition left inference.

For future work, we plan to lift our simulation result from
the ground case to the non-ground case.
This lifting will require the extension of the SCL calculus 
by an additional rule that learns clauses that are computed 
as intermediate steps during the conflict analysis. 
This rule was left out of previous versions of SCL because 
we would never use it in a CDCL inspired SCL-run and 
because it would have complicated the termination and 
non-redundancy proofs for SCL. (Although we are 
sure that the rule can be designed in such a way that all  
properties of the original calculus still hold.)
Considering the extension to the non-ground case,
this result can be used in various directions. It can be used
to develop an alternative implementation of the superposition calculus.
Given a fixed ordering, the trail can be developed according
to the ordering, generating only non-redundant superposition
inferences. On the other hand, the concept of finite saturation
can be kept this way preserving a strong mechanism for detecting
satisfiability. Secondly, the result means that SCL can be used
to naturally combine propagation driven reasoning with fixed ordering
driven reasoning. This might overcome some of the issues of the
current first-order portfolio approaches implemented in the state-of-the-art
provers.

\bibliographystyle{splncs04}

\begin{thebibliography}{10}
\providecommand{\url}[1]{\texttt{#1}}
\providecommand{\urlprefix}{URL }
\providecommand{\doi}[1]{https://doi.org/#1}

\bibitem{BachmairGanzinger94b}
Bachmair, L., Ganzinger, H.: Rewrite-based equational theorem proving with
  selection and simplification. Journal of Logic and Computation
  \textbf{4}(3),  217--247 (1994), revised version of Max-Planck-Institut
  f\"{u}r Informatik technical report, MPI-I-91-208, 1991

\bibitem{BachmairGanzinger01handbook}
Bachmair, L., Ganzinger, H.: Resolution theorem proving. In: Robinson, A.,
  Voronkov, A. (eds.) Handbook of Automated Reasoning, vol.~I, chap.~2, pp.
  19--99. Elsevier (2001)

\bibitem{BrombergerEtAl22}
Bromberger, M., Dragoste, I., Faqeh, R., Fetzer, C., Gonz{\'a}lez, L.,
  Kr{\"o}tzsch, M., Marx, M., Murali, H.K., Weidenbach, C.: A sorted datalog
  hammer for supervisor verification conditions modulo simple linear
  arithmetic. In: Fisman, D., Rosu, G. (eds.) Tools and Algorithms for the
  Construction and Analysis of Systems - 28th International Conference, {TACAS}
  2022, Held as Part of the European Joint Conferences on Theory and Practice
  of Software, {ETAPS}. Springer (2022)

\bibitem{BrombergerFW21}
Bromberger, M., Fiori, A., Weidenbach, C.: Deciding the bernays-schoenfinkel
  fragment over bounded difference constraints by simple clause learning over
  theories. In: Henglein, F., Shoham, S., Vizel, Y. (eds.) Verification, Model
  Checking, and Abstract Interpretation - 22nd International Conference,
  {VMCAI} 2021, Copenhagen, Denmark, January 17-19, 2021, Proceedings. Lecture
  Notes in Computer Science, vol. 12597, pp. 511--533. Springer (2021)

\bibitem{BrombergerGLW22}
Bromberger, M., Gehl, T., Leutgeb, L., Weidenbach, C.: A two-watched literal
  scheme for first-order logic. In: Konev, B., Schon, C., Steen, A. (eds.)
  Proceedings of the Workshop on Practical Aspects of Automated Reasoning
  Co-located with the 11th International Joint Conference on Automated
  Reasoning (FLoC/IJCAR 2022), Haifa, Israel, August, 11 - 12, 2022. {CEUR}
  Workshop Proceedings, vol.~3201. CEUR-WS.org (2022)

\bibitem{BrombergerEtAl22IJCAR}
Bromberger, M., Leutgeb, L., Weidenbach, C.: An efficient subsumption test
  pipeline for bs(lra) clauses. In: Blanchette, J., Kovacs, L., Pattinson, D.
  (eds.) Automated Reasoning - 11th International Joint Conference, {IJCAR}
  2022, Held as Part of the Federated Logic Conference, Proceedings. LNCS, vol.
  13385, pp. 147--168. Springer (2022)

\bibitem{BrombergerSW22}
Bromberger, M., Schwarz, S., Weidenbach, C.: Exploring partial models with
  {SCL}. In: Konev, B., Schon, C., Steen, A. (eds.) Proceedings of the Workshop
  on Practical Aspects of Automated Reasoning Co-located with the 11th
  International Joint Conference on Automated Reasoning (FLoC/IJCAR 2022),
  Haifa, Israel, August, 11 - 12, 2022. {CEUR} Workshop Proceedings, vol.~3201
  (2022)

\bibitem{BrombergerEtAl2023arxiv}
Bromberger, M., Schwarz, S., Weidenbach, C.: Scl(fol) revisited (2023).
  \doi{10.48550/ARXIV.2302.05954}, \url{https://arxiv.org/abs/2302.05954}

\bibitem{DeshernaisIsa23}
Desharnais, M.: {SCL}: Simple clause learning
  \url{https://bitbucket.org/isafol/isafol/src/master/Simple_Clause_Learning},
  Formal proof development

\bibitem{FioriWeidenbach19}
Fiori, A., Weidenbach, C.: Scl clause learning from simple models. In:
  Fontaine, P. (ed.) 27th International Conference on Automated Deduction,
  CADE-27. LNAI, vol. 11716. Springer (2019)

\bibitem{KnuthBendix70}
Knuth, D.E., Bendix, P.B.: Simple word problems in universal algebras. In:
  Leech, I. (ed.) Computational Problems in Abstract Algebra, pp. 263--297.
  Pergamon Press (1970)

\bibitem{Korovin08}
Korovin, K.: iprover - an instantiation-based theorem prover for first-order
  logic (system description). In: Armando, A., Baumgartner, P., Dowek, G.
  (eds.) Automated Reasoning, 4th International Joint Conference, {IJCAR} 2008,
  Sydney, Australia, August 12-15, 2008, Proceedings. Lecture Notes in Computer
  Science, vol.~5195, pp. 292--298. Springer (2008)

\bibitem{KovacsVoronkov13}
Kov{\'{a}}cs, L., Voronkov, A.: First-order theorem proving and vampire. In:
  Sharygina, N., Veith, H. (eds.) Computer Aided Verification - 25th
  International Conference, {CAV} 2013, Saint Petersburg, Russia, July 13-19,
  2013. Proceedings. Lecture Notes in Computer Science, vol.~8044, pp. 1--35.
  Springer (2013)

\bibitem{LeidingerWeidenbach22}
Leidinger, H., Weidenbach, C.: {SCL(EQ):} {SCL} for first-order logic with
  equality. In: Blanchette, J., Kov{\'{a}}cs, L., Pattinson, D. (eds.)
  Automated Reasoning - 11th International Joint Conference, {IJCAR} 2022,
  Haifa, Israel, August 8-10, 2022, Proceedings. Lecture Notes in Computer
  Science, vol. 13385, pp. 228--247. Springer (2022).
  \doi{10.1007/978-3-031-10769-6\_14}

\bibitem{NieuwenhuisRubio01handbook}
Nieuwenhuis, R., Rubio, A.: Paramodulation-based theorem proving. In: Robinson,
  A., Voronkov, A. (eds.) Handbook of Automated Reasoning, vol.~I, chap.~7, pp.
  371--443. Elsevier (2001)

\bibitem{SchlichtkrullEtAlIsa18}
Schlichtkrull, A., Blanchette, J.C., Traytel, D., Waldmann, U.: Formalization
  of bachmair and ganzinger's ordered resolution prover. Archive of Formal
  Proofs  (January 2018),
  \url{https://isa-afp.org/entries/Ordered_Resolution_Prover.html}, Formal
  proof development

\bibitem{SchulzEtAl19}
Schulz, S., Cruanes, S., Vukmirovi{\'c}, P.: Faster, higher, stronger: {E} 2.3.
  In: Fontaine, P. (ed.) Proc.\ of the 27th CADE, Natal, Brasil. pp. 495--507.
  No. 11716 in LNAI, Springer (2019)

\bibitem{WaldmannTRB20}
Waldmann, U., Tourret, S., Robillard, S., Blanchette, J.: A comprehensive
  framework for saturation theorem proving. In: Peltier, N.,
  Sofronie{-}Stokkermans, V. (eds.) Automated Reasoning - 10th International
  Joint Conference, {IJCAR} 2020, Paris, France, July 1-4, 2020, Proceedings,
  Part {I}. Lecture Notes in Computer Science, vol. 12166, pp. 316--334.
  Springer (2020). \doi{10.1007/978-3-030-51074-9\_18}

\bibitem{WeidenbachEtAlSpass2009}
Weidenbach, C., Dimova, D., Fietzke, A., Suda, M., Wischnewski, P.: Spass
  version 3.5. In: Schmidt, R.A. (ed.) 22nd International Conference on
  Automated Deduction (CADE-22). Lecture Notes in Artificial Intelligence,
  vol.~5663, pp. 140--145. Springer, Montreal, Canada (August 2009)

\end{thebibliography}

\section*{Appendix}

\subsection*{Lemma~\ref{lem:supstratapp} (SUP-MO Applicability)} 
  Let $N$ be a set of ground clauses.   
  If $N$ has a minimal false clause $C \neq \bot$, then there exists exactly one rule applicable to $N$ according to the SUP-MO strategy.
\begin{proof}
  For this proof, we make a case distinction:\newline
  Case 1: $C$ is the minimal false clause in $N$, and it has a maximal literal $L$ that is negative. 
          The strategy claims that in this case there 
          always exists a clause $D \in N$, with $D \prec C$, a strictly maximal literal $\comp(L)$, and $\delta_D = \{\comp(L)\}$.
          Note that $N_C \cup \delta_C \models L$ if $\comp(L) \not\in N_C \cup \delta_C$, 
          which also would imply that $N_C \cup \delta_C \models C$.
          Therefore, $\comp(L) \in N_C \cup \delta_C$.
          Moreover, we know that $\comp(L) \in N_C$ because the $\delta_C = \emptyset$ for every clause $C$ with a maximal literal that is negative.
          This means that, according to definition of the superposition model operator, 
          there exists a clause $D \in N$, with $D \prec C$, a strictly maximal literal $\comp(L)$, and $\delta_D = \{\comp(L)\}$.
          These properties also satisfy the conditions of the rule Superposition Left so the strategy rightfully claims that it is applicable.
  \newline        
  Case 2: $C$ is the minimal false clause in $N$, and it has a maximal literal $L$ that is positive. 
          The strategy claims that in this case $L$ is not strictly maximal in $C$ and therefore the rule Factoring is applicable.
          Assume to the contrary that $L$ is in fact strictly maximal, so $C = C' \lor L$ and $C' \prec \{L\}$.
          However, this also means that $N_C \not\models C'$ and, according to definition of the superposition model operator, 
          $\delta_C = \{L\}$ because $L$ is strictly maximal.
          Hence, $N_C \cup \delta_C \models C$, which is the contradiction we were aiming for.
\end{proof}

\subsection*{Lemma~\ref{lem:sclsupinisimul} (Initial SCL State Simulates Initial Superposition State)}
The initial SCL state $(\epsilon;N^0;\emptyset;\beta;0;\top)_{(0,\bot,\gamma)}$ simulates the initial superposition state $N^0$ and the model construction upto $N^0_{\bot} \cup \delta_{\bot}$
\begin{proof}
\begin{description}
    \item[(i) -- (iv)] Trivially true because the set of learned clauses is empty, the superposition state $N^0$ is the same as the input clause set, $\beta$ is sufficiently large by definition, and $\gamma$ is the identity function.
    \item[(v) -- (xiv)] Trivially true because both the trail and $N^0_{\bot} \cup \delta_{\bot}$ are empty, no atom is smaller than the minimal atom, and no clause is less or equal than $\bot$.
\end{description}
\end{proof}

\subsection*{Lemma~\ref{lem:sclsuppressimul} (SCL-SUP Preserves Simulation)}
Let the SCL state $S = (\Gamma;N^0;U;\beta;k;E)_{(i,C,\gamma)}$ simulate the superposition state $N^i$ and the corresponding model construction upto $N^i_{C'} \cup \delta_{C'}$,
where $C' = \gamma(C)$.
Let the SCL state $S' =(\Gamma';N^0;U';\beta;k';E')_{(j,D,\gamma')}$ be the result of one atomic sequence of SCL-SUP steps.
Then there exists a clause $D' \in N^j$ with $\gamma'(D) = D'$ 
and $S'$ simulates the superposition state $N^j$ and the model construction upto $N^j_{D'} \cup \delta_{D'}$.
\begin{proof}
We start with the 3 cases of atomic sequences according to Def.~\ref{def:sclsupstrat1}.
This means we can make the following additional assumptions due to Def.~\ref{def:sclsupstrat1}:
$E = \top$; $U' = U$; 
$D$ is the next largest clause from $C$ in the ordering $\prec_{\gamma}$ with respect to the ground clause set $N^0 \cup U$; 
$L$ is the maximal literal of $D$; 
$[\lnot A_1, \lnot A_2,\dots \lnot A_n]$ are all negative literals such that for all $i$ we have $A_i \prec L$, all $A_i$ undefined in $\Gamma$, $A_i$ occurs in $N^0 \cup U$, and  $A_i\prec A_{i+1}$.
Due to $S$ simulates $N^i$, 
we can assume that there exists a clause $D^* = \gamma(D)$ in $N^i$ such that $\sfac(D) = \sfac(D^*)$ (by Def.~\ref{def:sclsupstate}(ii)). 
Moreover, $j_0 + 1$ is the number of occurrences of $L$ in $D^*$.
\begin{description}
    \item[Case (1)$\rightarrow$(2a):] We additionally know by Def.~\ref{def:sclsupstrat1} that $L = B$ is a positive literal,
    $\Gamma'' = \Gamma, \neg A^{k+1}_1, \ldots, \neg A_n^{k+n}$,
    $\Gamma'' \not\models D$, 
    $\Gamma' = \Gamma,\neg A^{k+1}_1,\ldots, \neg A_n^{k+n}, B^{k+n+1}$, 
    $\gamma'$ is the same as $\gamma$ except that $\gamma'(D) = \sfac(D)$,
    $E' = \top$,
    $k' = k+n+1$,
    $j = i + j_0$,
    and Conflict is not applicable to $S'$.
	Due to $S$ simulates $N^i$, 
	we also know that $D$ is the smallest clause $D_2$ in $N^0 \cup U$ wrt.\ $\prec_{\gamma}$ such that $D_2 \models D$ 
	because otherwise $\Gamma \models D_2$ would hold (due to Def.~\ref{def:sclsupstate}(xii)), 
	and thus $\Gamma'' \models D$, which contradicts the assumption in step (2a) that $\Gamma'' \not\models D$.
	This also means that $D$ is the smallest clause $D_2$ in $N^0 \cup U$ wrt.\ $\prec_{\gamma}$ such that $\sfac(D_2) = \sfac(D)$.
	Analogously, there exists no clause $D^{**} \in N^i$ with $\sfac(D) = \sfac(D^{**})$ and less occurrences of $L$ than $D^*$
	because $D$ is the smallest clause in $N^0 \cup U$ wrt.\ $\prec_{\gamma}$ such that $D_2 \models D$
	and due to Def.~\ref{def:sclsupstate}(iv) $D^* = \gamma(D) \preceq D^{**}$ must hold.
    Moreover, for $m = 0, \ldots, j_0 - 1$ we know that the minimal false clause in $N^{i+m}$ is 
    $D^*_m = D_0 \lor L_1 \lor \ldots \lor L_{m}$,
    where $D^* = D_0 \lor L_1 \lor \ldots \lor L_{j_0}$, $L_1 = \ldots = L_{j_0} = L$,
    and $N^{i+m+1} = N^{i+m} \cup \{D^*_{m+1}\}$.
    This holds by the following inductive argument for $m$:
    (a)~$C' \prec D^*_m \prec D^*_{m-1} \prec \ldots \prec D^*_0$ by definition of $D^*_m$;
    (b)~$\delta_{D^*_m} = \emptyset$ because $L$ is not strictly maximal in $D^*_m$;
    (c)~$N^{i+m}_{D^*_m} = N^{i}_{C'} \cup \delta_{C'}$ due to (a) and (b);
    (d)~$\Gamma'' \not\models D^*_m$ because $\Gamma'' \not\models D$, 
    $D^*_m \models D$ as $\sfac(D) = \sfac(D^*_m)$,
    (e)~$N^{i+m}_{D^*_m} \cup \delta_{D^*_m} \not\models D^*_m$ because of (b),(c),(d), $S$ simulates $N^{i}_{C'} \cup \delta_{C'}$, 
    and none of the atoms $A^{k+1}_1, \ldots, A_n^{k+n}$ appear positively in $D^*_m$;
    (f) $D^*_0$ is the minimal false clause in $N^i$ because of (e) and Def.~\ref{def:sclsupstate}(iv),(xii);
    (g) if $D^*_m$ is the minimal false clause in $N^{i+m}$, then 
        SUP-MO applies Factoring to $D^*_m$ in $N^{i+m}$ which results in $N^{i+m+1} = N^{i+m} \cup\{D^*_{m+1}\}$;
    (h) $D^*_m$ is the minimal false clause in $N^{i+m}$ because of (a),(e),(f),(g), and by induction.
    (j) $D' = \sfac(D) \in N^j$ because of (g) and induction.
    \begin{description}
        \item[(i)]: $\atom(N^{i+m+1}) = \atom(N^0) = \atom(N^0 \cup U)$ holds because 
                    $\atom(N^i) = \atom(N^0) = \atom(N^0 \cup U)$ (since $S$ simulates $N^i$), 
                    $\atom(N^{i+m+1}) = \atom(N^{i+m}) \cup \atom(D^*_{m})$ for $m = 0, \ldots, j_0 - 1$, 
                    and $\atom(D^*_m) = \atom(D) \subseteq(\atom(N^0 \cup U))$ (since $\sfac(D) = \sfac(D^*_m)$);
                    $D \in N^0 \cup U$ due to assumptions derived from Def.~\ref{def:sclsupstrat1};
                    the rest follows directly from $S$ simulates $N^i$.
        \item[(ii)]: $\sfac(N^0 \cup U') \subseteq \sfac(N^j)$ holds because $U' = U$, $N^i \subseteq N^j$, and $\sfac(N^0 \cup U) \subseteq \sfac(N^i)$ 
                     (due to $S$ satisfies  Def.~\ref{def:sclsupstate}(ii)).
                     The rest holds for all $C^* \in (N^0 \cup U') \setminus D$ $\gamma'(C^*) \in N^j$ 
                     because $\gamma'(C^*) = \gamma(C^*)$ for $C^* \neq D$ and $S$ satisfies Def.~\ref{def:sclsupstate}(ii)).
                     $\gamma'(D) \in N^j$ and $\gamma'(D) = \sfac(D)$ for all $C \in (N^0 \cup U') \setminus D$
                     because $\gamma'(D) = \sfac(D)$ by definition Def.~\ref{def:sclsupstrat1} 
                     and $\sfac(D) \in N^j$ by argument (j) of the inductive subproof.
                     This also means that $\gamma'(D) = \sfac(D)$.
        \item[(iii)]: Holds for all $C^* \in N^i$ because $S$ satisfies Def.~\ref{def:sclsupstate}(iii) and $S'$ Def.~\ref{def:sclsupstate}(ii).
                      Holds for all $C^* \in N^j \setminus N^i$ because $\sfac(C^*) = \sfac(D^*) \in N^i$.
        \item[(iv)]: Holds for all $C^* \in N^i$ because $S$ satisfies Def.~\ref{def:sclsupstate}(iv), $S'$ Def.~\ref{def:sclsupstate}(ii), and $\gamma'(C^**) \preceq \gamma(C^**)$ for all clauses $C^**$.
                     Holds for all $C^* \in N^j \setminus N^i$ because $\sfac(C^*) = \sfac(D^*) = \sfac(D)$ and  $\gamma'(D)=\sfac(D^*)$.
        \item[(v),(viii),(ix)]: Holds for all atoms $A \in \Gamma$ and $A \in N^i_{C'} \cup \delta_{C'} = N^j_{C'} \cup \delta_{C'}$
                    because $S$ satisfies Def.~\ref{def:sclsupstate}(v),(viii),(ix), $\gamma'(C^**) \preceq \gamma(C^**)$ for all clauses $C^**$, 
                    and $S'$ satisfies Def.~\ref{def:sclsupstate}(ii), so if $\gamma(C^**) = \sfac(C^**)$, then $\gamma'(C^**) = \sfac(C^**)$.
                    The only other positive atom in $\Gamma'$ is $B$ and it is in $N^{j}_{D'} \cup \delta_{D'}$ 
                    because we can deduce from the above properties proven for the $D^*_m$ that 
                    $D^*_{j_0} \in N^j$,
                    $D' = \gamma(D^*) = \sfac(D^*) = D^*_{j_0}$,
                    $D^*_{j_0}$ is the next larger clause to $C'$, 
                    $D^*_{j_0}$ is not satisfied by $N^j_{C'} \cup \delta_{C'}$ , 
                    $B$ is the strictly maximal literal in $D^*_{j_0} = \sfac(D^*)$,
                    and thus $D^*_{j_0}$ produces $B$.
                    Hence, $B$ is the only atom in $N^j_{D'} \cup \delta_{D'} \setminus (N^j_{C'} \cup \delta_{C'})$.
        \item[(vi)]: We first prove that for all $A \in \atom(N^0)$ with $A \prec L$ and $\neg A \in \Gamma'$ it must 
                     hold that $A \not\in N^j_{D'} \cup \delta_{D'}$. 
                     This is the case because $N^j_{D'} \cup \delta_{D'}$ contains exactly all positive atoms in $\Gamma'$ 
                     (see the previous proof step for property (v)).
                     Next we prove that for all $A \in \atom(N^0)$ with $A \prec L$ and $A \not\in N^j_{D'} \cup \delta_{D'}$ it must 
                     hold that $\neg A \in \Gamma'$. 
                     Let $L'$ be the maximal literal in $C$ (and therefore also $C'$).
                     Then $\Gamma$ contains $\neg A$ for all literals $A \in \atom(N^0)$ with $A \prec L' \preceq L$ 
                     and $A \not\in N^i_{C'} \cup \delta_{C'} = N^j_{C'} \cup \delta_{C'}$ because $S$ satisfies Def.~\ref{def:sclsupstate}(vi).
                     The other $\neg A$ with $A \in \atom(N^0)$, $A \prec L$, and $A \not\in (N^j_{D'} \cup \delta_{D'})$
                     are in $\Gamma'$ due to the sequence of decisions $\neg A^{k+1}_1,\ldots, \neg A_n^{k+n}$
                     since this sequence includes all undefined atoms in $\Gamma$ with $A \in \atom(N^0)$ and $A \prec L$.
        \item[(vii)]: All atoms $A$ with $A \prec L'$, where $L'$ is the maximal literal in $C'$ are already
                      defined in $\Gamma$ (because $S$ satisfies Def.~\ref{def:sclsupstate}(v),(vi)), 
                      all atoms in $\Gamma$ are defined in ascending order of $\prec$ 
                      (because $S$ satisfies Def.~\ref{def:sclsupstate}(vii)), 
                      and the only atom $A$ with $L' \preceq A$ that can be defined in $\Gamma$ is the atom of $L'$.
                      Hence, all atoms in $\Gamma$ are ordered and contain no gaps.
                      By Def.~\ref{def:sclsupstrat1}, the atoms $\neg A^{k+1}_1,\ldots, \neg A_n^{k+n}, B^{k+n+1}$ 
                      appended to $\Gamma$ are therefore all larger than the atoms in $\Gamma$, 
                      ordered according to $\prec$, and contain no gaps.
                      Thus, the property also holds for $S'$.              
        \item[(x)-(xi)]: trivially satisfied because $\Gamma$ contains only decisions as it satisfies Def.~\ref{def:sclsupstate}(x) and 
                         SCL-SUP only adds decisions between $S$ and $S'$.
        \item[(xii)]: First of all, $\Gamma' \models D$ because $B$ is in $\Gamma'$.
                      Let $C^* \in N^0 \cup U$ be such that $\gamma'(C^*) \prec \gamma'(D)$.
                      Then $\gamma(C^*) \preceq \gamma(C)$ because $\gamma'(C^*) = \gamma(C^*)$ and $\gamma'(D) \preceq \gamma(D)$.
                      Therefore, $\Gamma \models C^*$ because $S$ satisfies Def.~\ref{def:sclsupstate}(xii),
                      which also means that $\Gamma' \models C^*$ because $\Gamma'$ is just an extension of $\Gamma$.              
        \item[(xiii)]: Conflict is not applicable to $S'$ by Def.~\ref{def:sclsupstrat1}.
        \item[(xiv)]: $\bot \not\in N^0 \cup U$ because $S$ satisfies Def.~\ref{def:sclsupstate}(xiv), 
                      $E' = \top \neq \bot$, and $\bot \not \in N^j$ because $S'$ satisfies Def.~\ref{def:sclsupstate}(iv).
    \end{description}
    \item[Case (1)$\rightarrow$(2b):] We additionally know by Def.~\ref{def:sclsupstrat1} that $L = B$ is a positive literal,
    $\Gamma'' = \Gamma, \neg A^{k+1}_1, \ldots, \neg A_n^{k+n}$,
    $\Gamma'' \not\models D$, 
    $\Gamma' = \Gamma,\neg A^{k+1}_1,\ldots, \neg A_n^{k+n}, B^{k+n+1}$,
    $\gamma'$ is the same as $\gamma$ except that $\gamma'(D) = \sfac(D)$, 
    $E'$ is the smallest clause in $N^0 \cup U$ wrt.\ $\prec_{\gamma}$ that is false wrt.\ $\Gamma,\neg A^{k+1}_1,\ldots, \neg A_n^{k+n},B^{\sfac(D)}$,
    $k' = k+n$, and
    $j = i + j_0$.
    Moreover, we can use the same deductions as in case (1)$\rightarrow$(2a) to show 
    properties (a)-(j) and for $m = 0, \ldots, j_0 - 1$ that the minimal false clause in $N^{i+m}$ is 
    $D^*_m = D_0 \lor L_1 \lor \ldots \lor L_{m}$,
    where $D^* = D_0 \lor L_1 \lor \ldots \lor L_{j_0}$, $L_1 = \ldots = L_{j_0} = L$,
    and $N^{i+m+1} = N^{i+m} \cup \{D^*_{m+1}\}$.
    \begin{description}
        \item[(i)-(ix),(xii),(xiv)]: Work the same as for case (1)$\rightarrow$(2a).           
        \item[(x),(xiii)]: trivially satisfied because $E' \not \in \{\top,\bot\}$.
        \item[(xi)]: $B^{\sfac(D)}$ is the topmost literal in $\Gamma'$ by Def.~\ref{def:sclsupstrat1};
                     $\Gamma, \neg A^{k+1}_1,\ldots, \neg A_n^{k+n}$ contains only decisions 
                     because $S$ satisfies Def.~\ref{def:sclsupstate}(x);
                     there must exist a clause $E^* \in N^{j}$ such that $\gamma(E') = E' = E^*$ 
                     because $S$ satisfies Def.~\ref{def:sclsupstate}(ii);
                     $\neg B$ is the largest literal in $E^*$ because $S'$ satisfies Def.~\ref{def:sclsupstate}(xii)
                     and so $\Gamma'$ satisfies all clauses with maximal literal $\preceq B$;
                     $N^{j}_{D'} \cup \delta_{D'} = N^{j}_{E^*} \cup \delta_{E^*}$ because 
                     $B \in N^{j}_{D'} \cup \delta_{D'}$ is the largest atom in $N^{j}_{D'} \cup \delta_{D'}$ 
                     and because $\neg B$ is the largest literal in $E^*$ the model operator does not produce any
                     other atoms between $D'$ and $E^*$;
                     $E^*$ is false in $N^{j}_{E^*} \cup \delta_{E^*}$ because 
                     $E'$ is false in $\Gamma'$, $E^* = E'$, and $S'$ satisfies Def.~\ref{def:sclsupstate}(v),(vi);
                     there is no smaller clause $E^*_2 \in N^{j}$ false in $N^{j}_{E^*} \cup \delta_{E^*}$ 
                     because $S'$ satisfies Def.~\ref{def:sclsupstate}(iv), so $N^0 \cup U$ would contain a clause 
                     $E'_2$ such that $E'_2 \models E^*_2$ and $\gamma(E'_2) \preceq E^*_2 \prec E'$ which contradicts our assumption 
                     that $E'$ is the smallest false clause in $\Gamma'$;
                     hence, $E'$  is the minimal false clause in $N^{j}$.
         \item[(xiv)]: $\bot \not\in N^0 \cup U$ because $S$ satisfies Def.~\ref{def:sclsupstate}(xiv), 
                       $E' \neq \bot$ because it has a maximal literal, 
                       and $\bot \not \in N^j$ because $S'$ satisfies Def.~\ref{def:sclsupstate}(iv).
    \end{description}
    \item[Case (1)$\rightarrow$(2c):] We additionally know by Def.~\ref{def:sclsupstrat1} that 
    $\Gamma' = \Gamma, \neg A^{k+1}_1, \ldots, \neg A_n^{k+n}$, 
    $\Gamma' \models D$,
    $\gamma' = \gamma$,
    $D' = \gamma(D)$,   
    $E' = \top$,
    $k' = k+n$, and
    $j = i$.
    \begin{description}
        \item[(i)-(iv),(vi),(vii),(xi),(xiv)]: Work the same as for case (1)$\rightarrow$(2a).           
        \item[(v),(viii),(ix)]: Holds for all atoms $A \in \Gamma$ and $A \in N^j_{C'} \cup \delta_{C'} = N^i_{C'} \cup \delta_{C'}$ 
                    because $S$ satisfies Def.~\ref{def:sclsupstate}(v),(viii),(ix) .
                    There is no further positive atom in $\Gamma'$ because SCL-SUP 
                    only adds negative literals between $S$ and $S'$ to the trail.
                    The other clauses $D^*$ between $C'$ and $D'$ 
                    also produce no positive atom in the superposition model construction
                    due to the following reasons.
                    The clause $D'$ is not productive in the superposition model construction
                    if its maximal literal is negative (see Def.~\ref{def:supmodelop}).
                    If the maximal literal of $D'$ is positive,
                    then $\Gamma' \models D'$ or one of the other two atomic sequences of Def.~\ref{def:sclsupstrat1} were applicable.
                    Since the atoms $A^{k+1}_1, \ldots, A_n^{k+n}$ appear if at all negatively in $D$ and thus $D'$ 
                    and since $S$ simulates $N^i_{C'} \cup \delta_{C'}$ and $C' \prec D'$,
                    it follows that $N^i_{D'} \models D'$.
                    Therefore, $D'$ is in case (1)$\rightarrow$(2c) never productive.
                    The other clauses $D^*$ between $C'$ and $D'$ (i.e. with $\gamma(C) = C' \prec D^* \prec D' = \gamma(D)$)
                    are also not productive because for every $D^* \in N^i$ 
                    there exists a $D_2 \in N^0 \cup U$ with $\gamma(D_2) \preceq \gamma(D^*)$ 
                    and $D_2 \models D^*$ (since $S$ satisfies Def.~\ref{def:sclsupstate}(iv)).
                    From $\gamma(D_2) \preceq \gamma(D^*)$ and $\gamma(C) = C' \prec D^* \prec D' = \gamma(D)$ follows 
                    that $\gamma(D_2) \preceq \gamma(C)$.
                    Therefore, $N^i_{D^*} \models D^*$ because 
                    $\Gamma \models D_2$ (since $S$ satisfies Def.~\ref{def:sclsupstate}(xii)) and
                    $S$ simulates $N^i_{C'} \cup \delta_{C'}$.
                    Hence, the other clauses $D^*$ between $C'$ and $D'$ are also not productive.
        \item[(x)]: Holds trivially because $E = \top$ and $\Gamma'$ contains only decided literals.      
        \item[(xii)]: $\Gamma' \models \Gamma$ because $\Gamma'$ is an extension of $\Gamma$;
                      $\Gamma \models C^*$ for all $C^* \in N^0 \cup U$ with $C^* \prec_{\gamma} D$ because 
                      $S$ satisfies Def.~\ref{def:sclsupstate}(xii) so 
                      $\Gamma' \models C^*$ for all $C^* \in N^0 \cup U$ with $C^* \prec C$;
                      $\Gamma' \models D$ because either the maximal literal $L$ of $D$ was negative
                      and therefore added to $\Gamma'$ by step (1) of the sequence or
                      the maximal literal $L$ of $D$ was positive and $\Gamma' \models D$ 
                      holds because otherwise $D$ would satisfy the preconditions of step (2a) or (2b).
        \item[(xiii)]: Proof by contradiction. 
                       Suppose there is a clause $E_2$ that is not false in $S$, but false in $S'$.
                       Then the maximal literal in $E_2$ is positive and one of the atoms 
                       $A_i$. However, this would also mean that $E_2$ is smaller than $D$ 
                       and therefore $E_2 \preceq C$.
                       This contradicts the fact that $S$ satisfies Def.~\ref{def:sclsupstate}(xii).
    \end{description}
\end{description}
We continue with the 4 cases of atomic sequences according to Def.~\ref{def:sclsupstrat2}.
This means we can make the following additional assumptions due to Def.~\ref{def:sclsupstrat2}:
$E \not\in \{\top,\bot\}$,
$L = \neg B$ is the maximal literal of $E$,
$\Gamma = \Gamma_0 B^{\sfac(C)}$,
$\Gamma_0$ contains only decision literals,
all literals in $\Gamma_0$ are smaller than $B$,
$\sfac(C) = C_1 \lor B$ with $C_1 \prec \{B\}$ because $B$ was propagated from $\sfac(C)$,
$E = E_0 \lor E_1$, where $E_1$ contains all occurrences of $L$ in $E$,
$j_0$ is the number of occurrences of $L$ in $E$,
and 
$E_2 = E_0 \lor C_1 \lor \ldots \lor C_{j_0}$,
where $C_1 = \ldots = C_{j_0}$.
Due to $S$ simulates $N^i$ and Def.~\ref{def:sclsupstate}(xi),
we can assume that 
$E \in N^i$, $\gamma(E) = E$, $E$ is the minimal false clause in $N^i$.
By Def.~\ref{def:sclsupstate}(iii),(viii),
there exists a clause $C' \in N^i$ with $C' = \sfac(C') = \sfac(C) = \gamma(C)$, $C' = C'_1 \lor B$, 
and $\delta_{C'} = \{B\}$.
Moreover, we can show for $m = 0, \ldots, j_0-1$ that $N^{i+m+1} = N^{i+m} \cup E^*_{m+1}$, 
that $E^*_{m}$ is the minimal false clause in $N^{i+m}$, 
and that $E^*_{m+1}$ is the result of applying Superposition Left to $E^*_{m}$ and $C'$,
where $E^*_m = E_0 \lor C_1 \lor \ldots \lor C_m \lor L_{m+1} \lor \ldots \lor L_{j_0}$,
$E = E_0 \lor L_1 \lor \ldots \lor L_{j_0}$, $C_1 = \ldots = C_{j_0}$, and $L_1 = \ldots = L_{j_0} = L$.
This holds because of induction:
(a2)~By induction hypothesis we can assume that $E^*_m$ is the minimal false clause in $N^{i}$;
(b2)~by Def.~\ref{def:supstrat}, the SUP-MO strategy reaches $N^{i+m+1}$ by applying Superposition Left to $E^*_m$ and $C'$, 
which adds the clause $E^*_{m+1}$ to $N^{i+m}$; 
(c2)~$E^*_{m+1} \prec E^*_{m}$ (for $j_0 > m \geq 0$) and $C' \prec E^*_{m}$ (for $j_0 > m \geq 0$) because $C_{m} \prec L_{m}$;
(d2)~for $m+1 < j_0$, $E^*_{m+1}$ is equivalent to $C_1 \lor E_0 \lor L$ modulo duplicate literals, $L$ is negative, $C_1 \prec L$, $E_0 \prec L$, and therefore $E^*_{m+1}$ is never productive;
(e2)~due to (c2) and (d2), $N^{i+m+1}_{E^*_{m+1}} \cup \delta_{E^*_{m+1}} = N^{i+m+1}_{C'} \cup \delta_{C'} = N^{i}_{C'} \cup \delta_{C'}$ for $m+1 < j_0$.
(f2)~due to (a2), (d2), and (e2), $E^*_{m+1}$ is false in $N^{i+m+1}_{E^*_{m+1}} \cup \delta_{E^*_{m+1}}$ for $m+1 < j_0$.
(g2)~due to (c2)and (f2), $E^*_{m+1}$ is also the minimal false clause in $N^{i+m+1}$;
(h2)~$E^*_{j_0} = E_2$.
\begin{description}
    \item[Case (1)$\rightarrow$(2a):] 
    We additionally know by Def.~\ref{def:sclsupstrat2} that 
    $\Gamma' = \epsilon$, $U' = U$, $\gamma' = \gamma$, $D = D' = \bot$, 
    $E' = E_2 = E_0 \lor C_1 \lor \ldots \lor C_1 = E^*_{j_0} = \bot$, 
    and therefore $C_1 = \bot$ and $E_2 = \bot$.
    \begin{description}
        \item[(i)]: Holds because $D = \bot$, $U' = U \cup \{E_2\}$, 
                    $N^j = N^i \cup \{E^*_1, \ldots, E^*_{j_0}\}$,
                    $\atom(E^*_1) = \ldots = \atom(E^*_{j_0-1}) = \atom(E^*_0) \cup \atom(C')$,
                    $\atom(E^*_{j_0}) = \atom(E_2) \subseteq \atom(E_0) \cup \atom(C)$,
                    $\atom(N^0) = \atom(N^i) = \atom(N^0 \cup U)$ because $S$ satisfies Def.~\ref{def:sclsupstate}(i),
                    $atom(E_0) \cup \atom(C) \cup \atom(E^*_0) \cup \atom(C') \subseteq \atom(N^0)$
                    because $E, C \in N^0 \cup U$, $E^*, C' \in N^i$ and $\atom(N^0) = \atom(N^i) = \atom(N^0 \cup U)$.
        \item[(ii)]: $\sfac(N^0 \cup U) \subseteq \sfac(N^i) \subseteq \sfac(N^j)$ because $S$ satisfies Def.~\ref{def:sclsupstate}(ii)
                     and $U = U'$.
                     The rest holds because $N^0 \cup U \subseteq N^i$, $\gamma' = \gamma$, and $S$ satisfies Def.~\ref{def:sclsupstate}(ii).                    
        \item[(iii)-(iv)]: Holds for all clauses in $N^i$ and $N^0 \cup U$ 
                           because $S$ satisfies Def.~\ref{def:sclsupstate}(iii),(iv) and $\gamma' = \gamma$. 
                           Holds for $\{E^*_1, \ldots, E^*_{j_0}\} = N^j \setminus N^{i}$
                           because $E' = E^*_{j_0} = \gamma(E')$, 
                           and for $m = 1, \ldots, j_0$ the maximum literal in $E^*_m$ is $L$ and negative, $E' \models E^*_m$, and $\gamma(E') = E' \preceq E^*_m$.
                           This holds because the clause $E^*_{m}$ has the form $E_0 \lor C_1 \lor \ldots \lor C_1 \lor L \lor \ldots \lor L$, 
                           $E_0 \prec L$ and $C_1 \prec L$,
                           $E'$ has the form $E_0 \lor C_1 \lor \ldots \lor C_1$, so
                           $E' \models E^*_m$ and $\gamma(E') = E' \preceq E^*_m$.
        \item[(v)-(ix)]: Holds trivially because $\Gamma' = \epsilon$ and $D = \bot$.
        \item[(x)-(xi),(xiii)]: Holds trivially because $E = \bot$ and $\Gamma' = \epsilon$.
        \item[(xii)]: Holds because $S$ satisfies Def.~\ref{def:sclsupstate}(xiv) and therefore $\bot \not\in N^0 \cup U$.
        \item[(xiv)]: $\bot \not\in N^0 \cup U$ because $S$ satisfies Def.~\ref{def:sclsupstate}(xiv); 
                      the rest holds because $E' = E^*_{j_0} = \bot$ and thus $\bot \in N^j$.
    \end{description}
    \item[Case (1)$\rightarrow$(4a):] 
    We additionally know by Def.~\ref{def:sclsupstrat2} that 
    $E' = E_2$, $U' = U \cup \{E_2\}$, $\gamma' = \gamma$,
    the maximal literal $L'_1 = \neg B_2$ of $E'$ is negative,
    $\Gamma_1$ is the prefix of $\Gamma_0$ that defines all atoms $A \prec B_2$,
    $D$ is the smallest clause in $N^0 \cup U$ wrt.\ $\prec_{\gamma}$ with maximum literal $B_2$ and $\Gamma_1 \not\models D$, 
    $\Gamma' = \Gamma_1, B_2^{\sfac(D)}$,
    $k'$ is the size of $\Gamma_1$,
    $\Gamma_1, B_2^{k'+1}$ is a prefix of $\Gamma_0$,
    and because $S$ satisfies Def.~\ref{def:sclsupstate}(viii),(ii) 
    there exists a $D' \in N^i \subseteq N^j$ with $\gamma(D) = \sfac(D) = \sfac(D') = D'$,
    and $\delta_{D'} = \{B\}$.
    Moreover, $N^j_{C^*} = N^i_{C^*}$ for all $C^* \preceq D'$ because $D' \prec \gamma(E_2)$.
    \begin{description}
        \item[(i)]: Holds because $U' = U \cup \{E_2\}$, $D = E_2 \in U'$, 
                    $N^j = N^i \cup \{E^*_1, \ldots, E^*_{j_0}\}$,
                    $\atom(E^*_1) = \ldots = \atom(E^*_{j_0-1}) = \atom(E^*_0) \cup \atom(C')$,
                    $\atom(E^*_{j_0}) = \atom(E_2) \subseteq \atom(E_0) \cup \atom(C)$,
                    $\atom(N^0) = \atom(N^i) = \atom(N^0 \cup U)$ because $S$ satisfies Def.~\ref{def:sclsupstate}(i),
                    $atom(E_0) \cup \atom(C) \cup \atom(E^*_0) \cup \atom(C') \subseteq \atom(N^0)$
                    because $E, C \in N^0 \cup U$, $E, C' \in N^i$ and $\atom(N^0) = \atom(N^i) = \atom(N^0 \cup U)$.
        \item[(ii)]: $\sfac(N^0 \cup U) \subseteq \sfac(N^i) \subseteq \sfac(N^j)$ because $S$ satisfies Def.~\ref{def:sclsupstate}(i).
                     $\sfac(E_2) \in \sfac(N^j)$ because $E^*_{j_0} = E_2 \in N^j$.
        \item[(iii)-(iv)]: Holds for all clauses in $N^i$ and $N^0 \cup U$ 
                           because $S$ satisfies Def.~\ref{def:sclsupstate}(iii),(iv) and $\gamma' = \gamma$. 
                           Holds for $\{E^*_1, \ldots, E^*_{j_0}\} = N^j \setminus N^{i}$ and 
                           $E_2$ due to (c2) and the definition of $E^*_{m}$ 
                           because $E_2 = E^*_{j_0}$,
                           the maximal literal in $E^*_{m}$ is negative for $m = 1, \ldots, j_0-1$, and 
                           $E_2 \models E^*_{m}$ and 
                           $E_2 \prec E^*_{m}$ for $m = 1, \ldots, j_0$.
        \item[(v)-(ix)]: Holds because $\gamma' = \gamma$,
                         $S$ satisfies Def.~\ref{def:sclsupstate}(v)-(ix),
                         $\Gamma'$ is a prefix of $\Gamma$ (except for the annotation of the last literal),
                         $\Gamma'$ defines by definition of the strategy all atoms $A \preceq B_2$,
                         and any producing clause $\preceq D'$ is in $N^i$ and 
                         any propagating clause $\preceq_{\gamma} D$ is in $N^0 \cup U$.
        \item[(x)]: Holds trivially because $E \not \in \{\top, \bot\}$ .
        \item[(xi)]: Holds because $\Gamma' = \Gamma_1, B_2^{\sfac(D)}$; $\Gamma'$ contains only decisions because 
                     it is a prefix of $\Gamma_0$, which contains only decisions;
                     $N^j$ contains $E^*_{j_0} = E_2 = \gamma(E_2)$;
                     and $E_2$ is the minimal false clause in $N^j$ because 
                     $E_2$ is the smallest clause added to $N^j$, 
                     $E_2 \prec E'$,
                     $E'$ was the minimal false clause in $N^i$ with respect to $\Gamma$, 
                     and $\Gamma'$ is only a prefix of $\Gamma$.
        \item[(xii)]: Holds because $S$ satisfies Def.~\ref{def:sclsupstate}(xii), $D \prec_{\gamma} C$, 
                      $\Gamma'$ is a prefix of $\Gamma$,
                      $\Gamma'$ defines by definition of the strategy all atoms $A \preceq B_2$,
                      Therefore, $\Gamma'$ always satisfies $D$.
                      $\Gamma'$ also satisfies all clauses $C^* \in N^0 \cup U'$ with $C^* \prec D$ because
                      all those clauses were already in $N^0 \cup U$, $\Gamma$ satisfied them 
                      (since $S$ satisfies Def.~\ref{def:sclsupstate}(xii)), 
                      and $\Gamma'$ is a prefix of $\Gamma$ that defines all atoms $\preceq B_2$.
        \item[(xiii)]: Conflict is not applicable because $E' \neq \top$.
        \item[(xiv)]: $\bot \not\in N^0 \cup U$ because $S$ satisfies Def.~\ref{def:sclsupstate}(xiv); 
                      the rest holds because $E_2 = E' \neq \bot$, $\bot \not \in N^i$, and $E^*_{m} \neq \bot$ (for $j_0 \geq m \geq 0$).
    \end{description}
    \item[Case (1)$\rightarrow$(4b):] 
    We additionally know by Def.~\ref{def:sclsupstrat2} that 
    $E' = \top$, $U' = U \cup \{E_2\}$, 
    the maximal literal $L'_1 = B_2$ of $E_2$ is positive
    and occurs $j_2+1$ times in $E_2$,
    $D = E_2$,
    $\gamma'$ is the same as $\gamma$ except that $\gamma'(E_2) = \sfac(E_2)$,
    $\Gamma_1$ is the prefix of $\Gamma_0$ that defines all atoms $A \prec L'_1$,
    $\Gamma' = \Gamma_1, B_2^{k'}$,
    $k'$ is the size of $\Gamma'$,
    $j_1 = i + j_0$,
    $j = j_1 + j_2$,
    and Conflict is not applicable to $S'$.    
    Moreover, we can use the same deductions as in case (1)$\rightarrow$(2a) for Def.~\ref{def:sclsupstrat1} to show 
    properties (a)-(j) and for $n = 0, \ldots, j_2 - 1$ that the minimal false clause in $N^{j_1+n}$ is 
    $D^*_n = D_0 \lor L'_1 \lor \ldots \lor L'_{n}$,
    where $E_2 = D_0 \lor L'_1 \lor \ldots \lor L'_{j_2}$, $L'_1 = \ldots = L'_{j_2}$,
    and $N^{j_1+n} = N^{j_1+n} \cup \{D^*_{n+1}\}$.
    Therefore, $D' = \sfac(E_2) = D^*_{j_2}$ and $D' \in N^j$.
    \begin{description}
    \item[(i)]: Note that $U' = U \cup \{E_2\}$, $D = E_2 \in U'$, 
                $N^j = N^i \cup \{E^*_1, \ldots, E^*_{j_0}\} 
                \cup \{D^*_1, \ldots, D^*_{j_2}\}$,
                $\atom(E^*_1) = \ldots = \atom(E^*_{j_0-1}) = \atom(E_0) \cup \atom(C)$,
                $\atom(E^*_{j_0}) = \atom(E_2) \subseteq \atom(E_0) \cup \atom(C)$,
                $\atom(D^*_{1}) = \ldots = \atom(D^*_{j_2}) = \atom(E_2)$,
                $\atom(N^0) = \atom(N^i) = \atom(N^0 \cup U)$ because $S$ satisfies Def.~\ref{def:sclsupstate}(i),
                $atom(E_0) \cup \atom(C) \subseteq \atom(N^0)$
                because $E, C \in N^0 \cup U$, $E, \sfac(C') \in N^i$ and $\atom(N^0) = \atom(N^i) = \atom(N^0 \cup U)$.
        \item[(ii)]: $\sfac(N^0 \cup U) \subseteq \sfac(N^i) \subseteq \sfac(N^j)$ because $S$ satisfies Def.~\ref{def:sclsupstate}(i).
                     $\sfac(E_2) \in \sfac(N^j)$ because $E^*_{j_0} = E_2 \in N^j$.
                     $\gamma'(E_2) = \sfac(E_2)$ and 
                     otherwise $\gamma'$ is identical to $\gamma$.
        \item[(iii)-(iv)]: Note that $N^j = N^i \cup \{E^*_1, \ldots, E^*_{j_0}\} 
                \cup \{D^*_1, \ldots, D^*_{j_2}\}$.
                           The properties hold for all clauses in $N^i$ and $N^0 \cup U$ 
                           because $S$ satisfies Def.~\ref{def:sclsupstate}(iii),(iv) and $\gamma' = \gamma$. 
                           The properties hold for $\{E^*_1, \ldots, E^*_{j_0}\}$ and 
                           $E_2$ due to (c2) and the definition of $E^*_{m}$ 
                           because $E_2 = E^*_{j_0}$,
                           the maximal literal in $E^*_{m}$ is negative for $m = 1, \ldots, j_0-1$, and 
                           $E_2 \models E^*_{m}$ and 
                           $E_2 \prec E^*_{m}$ for $m = 1, \ldots, j_0$.
                           The properties hold for $\{D^*_1, \ldots, D^*_{j_2}\}$ 
                           because $\gamma'(E_2) = \sfac(E_2) = D^*_n$,
                           for $n = 1, \ldots, j_2$, and 
                           therefore also $E_2 \models D^*_n$ and 
                           $\gamma'(E_2) \prec D^*_{n}$ for $n = 1, \ldots, j_2$.
        \item[(v)-(ix)]: Holds because
                         $S$ satisfies Def.~\ref{def:sclsupstate}(v)-(ix),
                         $\Gamma_1$ is a prefix of $\Gamma$,
                         $\Gamma_1$ defines by definition of the strategy all atoms $A \prec L'_1$,
                         therefore any producing clause $\prec D' = \sfac(E_2)$ is already in $N^i$ and 
                         any propagating clause $\prec_{\gamma'} D$ in $N^0 \cup U$.
                         The only additional atom in $\Gamma' = \Gamma_1, B_2^{k'}$ 
                         and $N^j_{D'} \cup \delta_{D'}$ is $B_2$ 
                         and it is produced by $D'$ and can be propagated by $D$.
                         Moreover, $D' = \sfac(D') = \sfac(D) = \gamma'(D)$.  
        \item[(x)]: Holds because $\Gamma' = \Gamma_1, B_2^{k'}$, 
                    $\Gamma_1$ is a prefix of $\Gamma_0$, $S$ satisfies Def.~\ref{def:sclsupstate}(xi), 
                    and therefore $\Gamma_0$ and $\Gamma_1$ only contain decisions.
        \item[(xi)]: Holds trivially because $E = \top$.
        \item[(xii)]: Holds for $D = E_2$ because $B_2$ is true in $\Gamma'$.
                      Holds for all clauses $C^* \in N^0 \cup U'$ with $C^* \prec_{\gamma'} D$ 
                      because $S$ satisfies Def.~\ref{def:sclsupstate}(xii) and so $\Gamma \models C^*$,  
                      $\Gamma_1, \comp(B_2)$ is a prefix of $\Gamma$,
                      $\Gamma_1, \comp(B_2)$ defines by definition of the strategy all atoms $A \preceq B_2$,
                      the maximal literal in $C^*$ is $A \preceq B_2$, 
                      so $C^*$ is true independent of $B_2$, 
                      which means $\Gamma_1 \models C^*$ and therefore  $\Gamma_1, B \models C^*$.
        \item[(xiii)]: Conflict is not applicable due to the precondition of Def.~\ref{def:sclsupstrat2}(4b).
        \item[(xiv)]: $\bot \not\in N^0 \cup U$ because $S$ satisfies Def.~\ref{def:sclsupstate}(xiv); 
                      the rest holds because $E' \neq \bot$, $\bot \not \in N^i$, and $E^*_{m} \neq \bot$ (for $j_0 \geq m \geq 0$).
    \end{description}
    \item[Case (1)$\rightarrow$(4c):]
    We additionally know by Def.~\ref{def:sclsupstrat2} that 
    $U' = U \cup \{E_2\}$, 
    the maximal literal $L'_1 = B_2$ of $E_2$ is positive
    and occurs $j_2+1$ times in $E_2$,
    $D = E_2$,
    $\gamma'$ is the same as $\gamma$ except that $\gamma'(E_2) = \sfac(E_2)$,
    $\Gamma_1$ is the prefix of $\Gamma_0$ that defines all atoms $A \prec L'_1$,
    $\Gamma' = \Gamma_1, B_2^{\sfac(D)}$,
    $k'$ is the size of $\Gamma'$,
    $j_1 = i + j_0$,
    $j = j_1 + j_2$,
    $E'$ is the minimal clause that is false with respect to $\Gamma'$,
    and $\gamma'(E') = E'$ because the maximal literal in $E'$ is negative.     
    Moreover, we can use the same deductions as in case (1)$\rightarrow$(2a) for Def.~\ref{def:sclsupstrat1} to show 
    properties (a)-(j) and for $n = 0, \ldots, j_2 - 1$ that the minimal false clause in $N^{j_1+n}$ is 
    $D^*_n = D_0 \lor L'_1 \lor \ldots \lor L'_{n}$,
    where $E_2 = D_0 \lor L'_1 \lor \ldots \lor L'_{j_2}$, $L'_1 = \ldots = L'_{j_2}$,
    and $N^{j_1+n} = N^{j_1+n} \cup \{D^*_{n+1}\}$.
    Therefore, $D' = \sfac(E_2) = D^*_{j_2}$ and $D' \in N^j$.
    \begin{description}
        \item[(i)-(ix),(xii),(xiv)]: Work the same as for case (1)$\rightarrow$(4b).
        \item[(x)]: Holds trivially because $E \neq \top$.
        \item[(xi)]: Holds because $\Gamma' = \Gamma_1, B_2^{\sfac(D)}$, 
                     $\Gamma_1$ is a prefix of $\Gamma_0$, $S$ satisfies Def.~\ref{def:sclsupstate}(xi), 
                     and therefore $\Gamma_0$ and $\Gamma_1$ only contain decisions.  
                     $D^*_{j_2} = D' = \sfac(D') = \sfac(D) = \gamma'(D)$ due to properties we have proven for $D^*_n$;
                     $\Gamma_1$ contains only decisions because $\Gamma_1$ is a prefix of $\Gamma_0$ and $S$ satisfies Def.~\ref{def:sclsupstate}(x);
                     it must hold that $E' \in N^{j}$  
                     because $\gamma'(E') = E'$ and $S'$ satisfies Def.~\ref{def:sclsupstate}(ii);
                     $\neg B$ is the largest literal in $E'$ because $S'$ satisfies Def.~\ref{def:sclsupstate}(xii)
                     and so $\Gamma'$ satisfies all clauses with maximal literal $\preceq B$;
                     $N^{j}_{D'} \cup \delta_{D'} = N^{j}_{E'} \cup \delta_{E'}$ because 
                     $B_2 \in N^{j}_{D'} \cup \delta_{D'}$ is the largest atom in $N^{j}_{D'} \cup \delta_{D'}$ 
                     and because $\neg B$ is the largest literal in $E'$ the model operator does not produce any
                     other atoms between $D'$ and $E'$;
                     $E'$ is false in $N^{j}_{E'} \cup \delta_{E'}$ because 
                     $E'$ is false in $\Gamma'$, and $S'$ satisfies Def.~\ref{def:sclsupstate}(v),(vi);
                     there is no smaller clause $E^{**} \in N^{j}$ false in $N^{j}_{E'} \cup \delta_{E'}$ 
                     because $S'$ satisfies Def.~\ref{def:sclsupstate}(iv), so $N^0 \cup U'$ would contain a clause 
                     $E''$ such that $E'' \models E^{**}$ and $\gamma(E'') \preceq E^{**} \prec E'$ which contradicts our assumption 
                     that $E'$ is the smallest false clause in $\Gamma'$;
                     hence, $E'$  is the minimal false clause in $N^{j}$.
        \item[(xiii)]: Conflict is not applicable because $E \neq \top$.
    \end{description} 
\end{description}
\end{proof}

\subsection*{Lemma~\ref{lem:sclsupadvances} (SCL-SUP Advances the Simulation) }
Let the SCL state $S = (\Gamma;N^0;U;\beta;k;E)_{(i,D,\gamma)}$ simulate the superposition state $N^i$ and the model construction upto $N^i_{\gamma(D)} \cup \delta_{\gamma(D)}$.
Let the SCL state $S' =(\Gamma';N^0;U';\beta;k';E')_{(j,D',\gamma')}$ be the next state reachable by one atomic sequence of SCL-SUP steps.
Then either $i < j$ or $i = j$ and $\gamma' = \gamma$ and $D \prec_{\gamma} D'$.
\begin{proof}
If the atomic sequence is according to Def.~\ref{def:sclsupstrat1}, then $D'$ is the next larger clause wrt.\ $\prec_{\gamma}$ and $D$.
If $\gamma(D') = \sfac(D')$, then $i = j$, $\gamma' = \gamma$, and $D \prec_{\gamma} D'$ hold.
Otherwise, SCL-SUP skips some Factoring steps, so $i < j$.
If the atomic sequence is according to Def.~\ref{def:sclsupstrat2}, then by definition the annotated number increases by one for every resolution step in the sequence. Since there is at most one resolution step, $i < j$ always holds.
\end{proof}

\subsection*{Lemma~\ref{lem:sclsupfinal} (SCL-SUP Correctness of Final States)}
Let the SCL state $S = (\Gamma;N^0;U;\beta;k;E)_{(i,D,\gamma)}$ simulate the superposition state $N^i$ and the model construction upto $N^i_{\gamma(D)} \cup \delta_{\gamma(D)}$.
Let there be no more states reachable from $S$ following an atomic sequence of SCL-SUP steps.
Then $S$ is a \emph{final state}, i.e., either (i)~$E = \bot$, $D = \bot$, $\bot \in N^i$, and $N^0$ is unsatisfiable or 
(ii)~$\Gamma \models N^0$.
\begin{proof}
Proof by case distinction:
\begin{description}
  \item[$E = \bot$]: Then by property Def.~\ref{def:sclsupstate}(xiv) $\bot \in N^i$ and $\Gamma = \epsilon$.
  Since superposition is sound this means that $N^0$ is unsatisfiable. 
  Moreover, $D = \bot$ or due to Def.~\ref{def:sclsupstate}(xii) the empty trail would satisfy a non-empty clause, which would be a contradiction. Therefore, this SCL state is a final state.
  \item[$E = \top$ and $D$ is the largest clause in $N^0 \cup U$ wrt.\ $\prec_{\gamma}$]: 
  $\Gamma \models N^0$ because of properties Def.~\ref{def:sclsupstate}(xii) and (iv).
  Therefore, this SCL state is a final state.
  \item[$E = \top$ and $D$ is not the largest clause in $N^0 \cup U$ wrt.\ $\prec_{\gamma}$]: 
  Then an atomic sequence of SCL-SUP steps as described in Def.~\ref{def:sclsupstrat1} must fulfill the following preconditions:
  (a)~$E = \top$ and there exists a next largest clause in $N^0 \cup U$ wrt.\ $\prec_{\gamma}$ and $D$, 
      which is automatically fulfilled by the assumptions for the given case; and
  (b)~$N^i$ contains $D' = \gamma(D)$ and $\sfac(D) = \sfac(D')$,
  which is fulfilled because $S$ satisfies property Def.~\ref{def:sclsupstate}(ii).
  \item[$E \not\in \{\bot,\top\}$]: Then by property Def.~\ref{def:sclsupstate}(xi) $\Gamma = \Gamma', B^{\sfac(D)}$, 
  $\Gamma'$ contains only decisions, 
  and there exists an $E' \in N^i$  that is the minimal false clause in $N^i$ such that $\gamma(E) = E = E'$ and $\neg B \in E$.
  By properties Def.~\ref{def:sclsupstate}(v)-(vii), all atoms $A \preceq B$ are defined in $\Gamma$ in ascending order $\prec$.
  By property Def.~\ref{def:sclsupstate}(viii), there exists a clause $D' \in N^i$ such that $\sfac(D') = D'$, $\gamma(D) = \sfac(D) = D'$, 
  and $\delta_{D'} = \{B\}$.
  This satisfies all non-trivial preconditions of Def.~\ref{def:sclsupstrat2}, 
  so the atomic sequence of SCL-SUP steps described in Def.~\ref{def:sclsupstrat2} is applicable.
\end{description}
\end{proof}

\subsection*{Lemma~\ref{lem:sclsupreasstrat} (SCL-SUP is a Regular SCL Strategy)}
SCL-SUP is a regular SCL strategy if it is executed on a state 
$S = (\Gamma;N^0;U;\beta;k;E)_{(i,C,\gamma)}$ that simulates a superposition state $N^i$ and the corresponding model construction upto $N^i_{\gamma(C)} \cup \delta_{\gamma(C)}$.
\begin{proof}
Let $S' =(\Gamma';N^0;U';\beta;k';E')_{(j,D',\gamma')}$ be the next state reachable by one atomic sequence of SCL-SUP steps from $S$.
Let $L$ be the largest literal in $D'$.
First note that SCL-SUP only applies the rule Conflict directly after an application of the rule Propagate (see steps Def.~\ref{def:sclsupstrat1}(2b), Def.~\ref{def:sclsupstrat2}(4a), and Def.~\ref{def:sclsupstrat2}(4c)).
Therefore, SCL-SUP is a regular SCL strategy if it applies the rule Conflict within the atomic sequence of steps as soon as Conflict is applicable.
We prove this by a case distinction:\newline
Case 1: The atomic sequence follows Def.~\ref{def:sclsupstrat1}.
This means $\Gamma'$ is an extension of $\Gamma$, $U = U'$, and the only rules applied are Decide, Propagate, and Conflict.
Furthermore, due to Lemma~\ref{lem:sclsuppressimul}, 
we know that $\Gamma'$ defines all atoms $A$ with $A \prec L$ (Def.~\ref{def:sclsupstate}(v)-(vii)), 
$\Gamma'$ defines none of the atoms $A$ with $L \prec A$ (Def.~\ref{def:sclsupstate}(v)-(vii)),
$A$ or $\comp(A)$ is the last literal in $\Gamma'$ if $\Gamma'$ defines $A$ (Def.~\ref{def:sclsupstate}(v)-(vii)), and 
$\Gamma$ satisfies all clauses with maximum literal $L' \prec L$ (Def.~\ref{def:sclsupstate}(xii)).
Moreover, every clause in $N^0 \cup U$ is not false with respect to $\Gamma'$ if it has a maximum literal $L'$ such that $L \prec \atom(L')$ because at least $L'$ is not yet defined.
This also means, that such a clause is not false with respect to any prefix of $\Gamma'$.
Therefore, the only clauses in $N^0 \cup U$ that can be false with respect to $\Gamma'$ have maximum literal $L$ or $\comp(L)$ and 
they are not false in a strict prefix of $\Gamma'$ as $L$ is not yet defined in a strict prefix of $\Gamma'$.
Hence, Conflict is only applicable after $L$ is defined, which is also the case when SCL-SUP checks for a false clause and applies Conflict.\newline
Case 2: The atomic sequence follows Def.~\ref{def:sclsupstrat2} and ends in case (2a).
Then Conflict was never applicable between $S$ and $S'$ because the last argument in the SCL state was always $\neq \top$.\newline
Case 3: The atomic sequence follows Def.~\ref{def:sclsupstrat2} and ends in case (4a).
Then Conflict was not applicable between steps (1) and (3) of the strategy because the last argument in the SCL state was always $\neq \top$.
In part (4a) of the strategy only rules Propagate and Conflict are applied.
Moreover, $\Gamma'$ is a strict prefix of $\Gamma = \Gamma_0, B$ (except for the annotation of the last literal on the trail), 
$U' = U \cup \{E'\}$, $B$ is the top literal in $C$, 
$\comp(B)$ is the top literal in $E$, and the top literal $L_1$ in $E'$ is smaller than $B$ (as any occurrence $\comp(B)$ was resolved away).
Similar to what we have shown for case 1, 
we know that $\Gamma$ defines all atoms $A$ with $A \prec B$ (Def.~\ref{def:sclsupstate}(v)-(vii)), 
$\Gamma$ defines none of the atoms $A$ with $B \prec A$ (Def.~\ref{def:sclsupstate}(v)-(vii)),
$B$ is the last literal in $\Gamma$ (Def.~\ref{def:sclsupstate}(v)-(vii)), and 
$\Gamma$ satisfies all clauses with maximum literal $L' \prec B$ (Def.~\ref{def:sclsupstate}(xii)).
Moreover, every clause in $N^0 \cup U$ is not false with respect to $\Gamma$ if it has a maximum literal $L'$ such that $B \prec \atom(L')$ because at least $L'$ is not yet defined.
This also means, that such a clause is not false with respect to any prefix of $\Gamma$.
Therefore, the only clauses in $N^0 \cup U$ that can be false with respect to $\Gamma$ have maximum literal $L=B$ or $\comp(L)$ and 
there are no false clauses in $N^0 \cup U$ with respect to the strict prefix $\Gamma'$ of $\Gamma$ as $L$ is not yet defined in $\Gamma'$.
Thus, the only clause that can be false with respect to $\Gamma'$ is the newly learned clause $E'$ and it can only be false after its maximal literal has been defined.
Hence, Conflict is only applicable after $L_1$ is defined, which is also the case when SCL-SUP applies Conflict.\newline
Case 4: The atomic sequence follows Def.~\ref{def:sclsupstrat2} and ends in case (4b).
Then Conflict was not applicable between steps (1) and (3) of the strategy because the last argument in the SCL state was always $\neq \top$.
In part (4b) of the strategy only Decide is applied.
Furthermore, due to Def.~\ref{def:sclsupstate}(xiii), Conflict is not applicable in state $S'$ and therefore
also not in any state $S^* = (\Gamma^*;N^0;U';\beta;k';\top)$ visited in part (4a) of the strategy because 
$S^*$ is equivalent to $S'$ except that $\Gamma^*$ is a prefix of $\Gamma'$.\newline
Case 5: The atomic sequence follows Def.~\ref{def:sclsupstrat2} and ends in case (4c).
Then Conflict was not applicable between steps (1) and (3) of the strategy because the last argument in the SCL state was always $\neq \top$.
In part (4c) of the strategy only Propagate and Conflict are applied and $\Gamma' = \Gamma_1, B_2^{\sfac(D')}$, where $\Gamma_1$ is a prefix of $\Gamma_0$ and $\Gamma = \Gamma_0, B$ and $B_2 \prec B$.
Furthermore, due to Lemma~\ref{lem:sclsuppressimul}, 
we know that $\Gamma_1$ defines all atoms $A$ with $A \prec B_2$ (Def.~\ref{def:sclsupstate}(v)-(vii)), 
$\Gamma_1$ defines none of the atoms $A$ with $B_2 \preceq A$ (Def.~\ref{def:sclsupstate}(v)-(vii)), 
$\Gamma$ satisfies all clauses in $N^i$ with maximum literal $L' \preceq B$ (Def.~\ref{def:sclsupstate}(xii))
and therefore $\Gamma_1$ falsifies no of the clauses in $N^i$, and 
$\Gamma'$ falsifies none of the clauses in $N^j \setminus N^i$ because they all contain the literal $B_2$ (see proof for Lemma~\ref{lem:sclsuppressimul})
and therefore $\Gamma_1$ falsifies no of the clauses in $N^j \setminus N^i$.
Therefore, Conflict is not applicable before $B_2$ is propagated.
\end{proof}

\end{document}